\documentclass[letterpaper,10pt,journal,twoside]{IEEEtran}
\IEEEoverridecommandlockouts
\usepackage{amsmath,amssymb,amsfonts}
\usepackage{graphicx}
\usepackage{textcomp}
\usepackage{xcolor}
\usepackage{threeparttable}
\usepackage{tabularx}
\usepackage{booktabs}
\usepackage{multirow}
\usepackage{colortbl}
\usepackage{enumitem}
\usepackage{subfigure}
\usepackage{pifont}
\usepackage{comment}
\usepackage{array} 

\usepackage{bigstrut,multirow,rotating}
\usepackage[numbers,sort&compress]{natbib}
\usepackage{tikz}
\usepackage{threeparttable}
\usepackage[colorlinks, linkcolor=magenta, anchorcolor=green, citecolor=blue]{hyperref}

\newcommand{\hl}{\textcolor{black}}
\newcommand{\hlblue}{\textcolor{black}}
\newcommand{\hblue}{\textcolor{black}}
\usepackage[normalem]{ulem}   
\usepackage{amsmath}

\usepackage{algorithm}
\usepackage{algorithmicx}
\usepackage{algpseudocode}
\usepackage{amsmath}
\usepackage[rightcaption]{sidecap}

\def\BibTeX{{\rm B\kern-.05em{\sc i\kern-.025em b}\kern-.08em
    T\kern-.1667em\lower.7ex\hbox{E}\kern-.125emX}}
    
\newcommand*{\circled}[1]{\lower.7ex\hbox{\tikz\draw (0pt, 0pt)%
    circle (.5em) node {\makebox[1em][c]{\small #1}};}}

\usepackage{titlesec}
\titlespacing\section{0pt}{3pt plus 1pt minus 1pt}{0pt plus 1pt minus 1pt}
\titlespacing\subsection{0pt}{3pt plus 1pt minus 1pt}{0pt plus 1pt minus 1pt}
\titlespacing\subsubsection{0pt}{3pt plus 1pt minus 1pt}{2pt plus 1pt minus 1pt}
\begin{document}


\title{\huge
HPIM: \underline{H}eterogeneous \underline{P}rocessing-\underline{I}n-\underline{M}emory-based Accelerator for Large Language Models Inference
}

\author{Cenlin~Duan,
        Jianlei~Yang,~\IEEEmembership{Senior Member,~IEEE,}
        Rubing~Yang,
        Yikun~Wang,
        Yiou~Wang,
        Lingkun~Long,
        Yingjie~Qi,
        Xiaolin~He,
        Ao~Zhou,
        Xueyan~Wang,~\IEEEmembership{Member,~IEEE,}
        and~Weisheng~Zhao,~\IEEEmembership{Fellow,~IEEE}
\thanks{Manuscript received in July 2025, revised in January 2026 and April 2026, accepted in June 2026. This work is supported by the National Natural Science Foundation of China (Grant No. 62572036). 
\textit{Corresponding authors are Jianlei Yang and Weisheng Zhao.}
}
\thanks{C. Duan and W. Zhao are with the School of Integrated Circuit Science and Engineering, Beihang University, Beijing, 100191, China, and the Shandong Laboratory of ASIC and Intelligent Instruments in Qingdao, Beihang University, Qingdao 266400, China. E-mail: \url{weisheng.zhao@buaa.edu.cn}}
\thanks{J. Yang, R. Yang, Y. Wang, Y. Wang, L. Long, Y. Qi, X. He and A. Zhou are with the School of Computer Science and Engineering, Beihang University, Beijing, 100191, China. E-mail: \url{jianlei@buaa.edu.cn}
}
\thanks{X. Wang is with the School of Integrated Circuit Science and Engineering, Beihang University, Beijing, 100191, China.}

}

\maketitle
\bstctlcite{IEEEexample:BSTcontrol}

\begin{abstract}

The deployment of large language models (LLMs) presents significant challenges due to their enormous memory footprints, low arithmetic intensity, and stringent latency requirements, particularly during the autoregressive decoding stage.
Traditional compute-centric accelerators, such as GPUs, suffer from severe resource underutilization and memory bandwidth bottlenecks in these memory-bound workloads. 
To overcome these fundamental limitations, we propose HPIM, the first memory-centric heterogeneous Processing-In-Memory (PIM) accelerator that integrates SRAM-PIM and HBM-PIM subsystems designed specifically for LLM inference.
HPIM employs a software-hardware co-design approach that combines a specialized compiler framework with a heterogeneous hardware architecture. 
It intelligently partitions workloads based on their characteristics: latency-critical attention operations are mapped to the SRAM-PIM subsystem to exploit its ultra-low latency and high computational flexibility, while weight-intensive GEMV computations are assigned to the HBM-PIM subsystem to leverage its high internal bandwidth and large storage capacity.
Furthermore, HPIM introduces a tightly coupled pipeline strategy across SRAM-PIM and HBM-PIM subsystems to maximize intra-token parallelism, thereby significantly mitigating the serial dependency of the autoregressive decoding stage.
Comprehensive evaluations using a cycle-accurate simulator demonstrate that HPIM significantly outperforms state-of-the-art accelerators, achieving a peak speedup of up to $23.1\times$ compared to the NVIDIA A100 GPU. 
Moreover, HPIM exhibits superior performance over contemporary PIM-based accelerators, highlighting its potential as a highly practical and scalable solution for accelerating large-scale LLM inference.

\end{abstract}

\begin{IEEEkeywords}
Large Language Models, Heterogeneous Processing-In-Memory, HBM-PIM, SRAM-PIM
\end{IEEEkeywords}

\IEEEpeerreviewmaketitle

\section{Introduction}\label{sec:Introduction}

Large Language Models (LLMs), such as the GPT family~\cite{brown2020language, radford2019language} and the LLaMA family~\cite{grattafiori2024llama, touvron2023llama2}, 
have dramatically transformed the field of artificial intelligence, enabling unprecedented capabilities in conversational agents~\cite{patil2024gorilla}, text generation~\cite{dao2022flashattention}, and code synthesis~\cite{zhang2024autocoderover}.
However, the deployment of these models is fundamentally limited by their inherently low computational intensity, particularly during the autoregressive decoding phase.
This memory-bound characteristic imposes a severe bandwidth bottleneck on conventional accelerators, such as GPUs, which are optimized for compute-intensive workloads.
Such a fundamental mismatch directly leads to severe compute underutilization, degrades overall throughput and efficiency, and ultimately motivates a paradigm shift towards novel architectural solutions.

\begin{figure}[t]
\centering
\includegraphics[width =\linewidth]{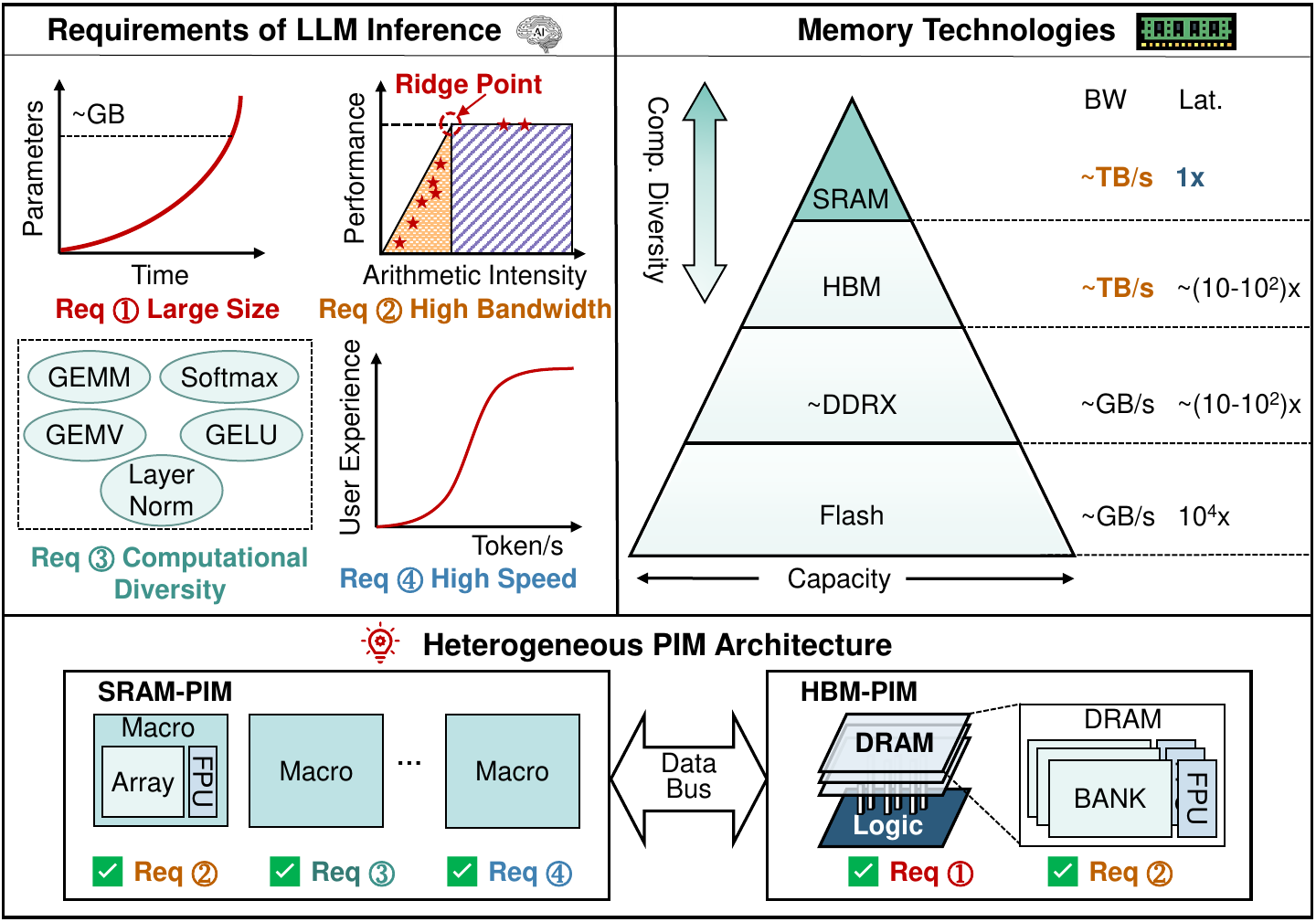}
\vspace{-18pt}
\caption{The diverse requirements of LLM inference, the corresponding trade-offs in the "PIM pyramid", and the advantages of a heterogeneous PIM architecture.}
\label{fig1:PIM pyramid}
\end{figure}
To address these challenges, Processing-In-Memory (PIM)~\cite{duan2023ddc, duan2024towards, qi2025cimflow, he2026miredo} has emerged as a promising architectural paradigm. 
PIM aims to mitigate the data movement bottleneck by integrating computational logic directly within or near memory arrays to exploit massive internal memory bandwidth.
It can be implemented across a spectrum of memory technologies, each presenting a distinct trade-off between capacity, bandwidth, computational diversity, and latency, as illustrated by the "PIM pyramid" in Fig.~\ref{fig1:PIM pyramid}.
\begin{figure*}[t]
\centering
\includegraphics[width =0.8\linewidth]{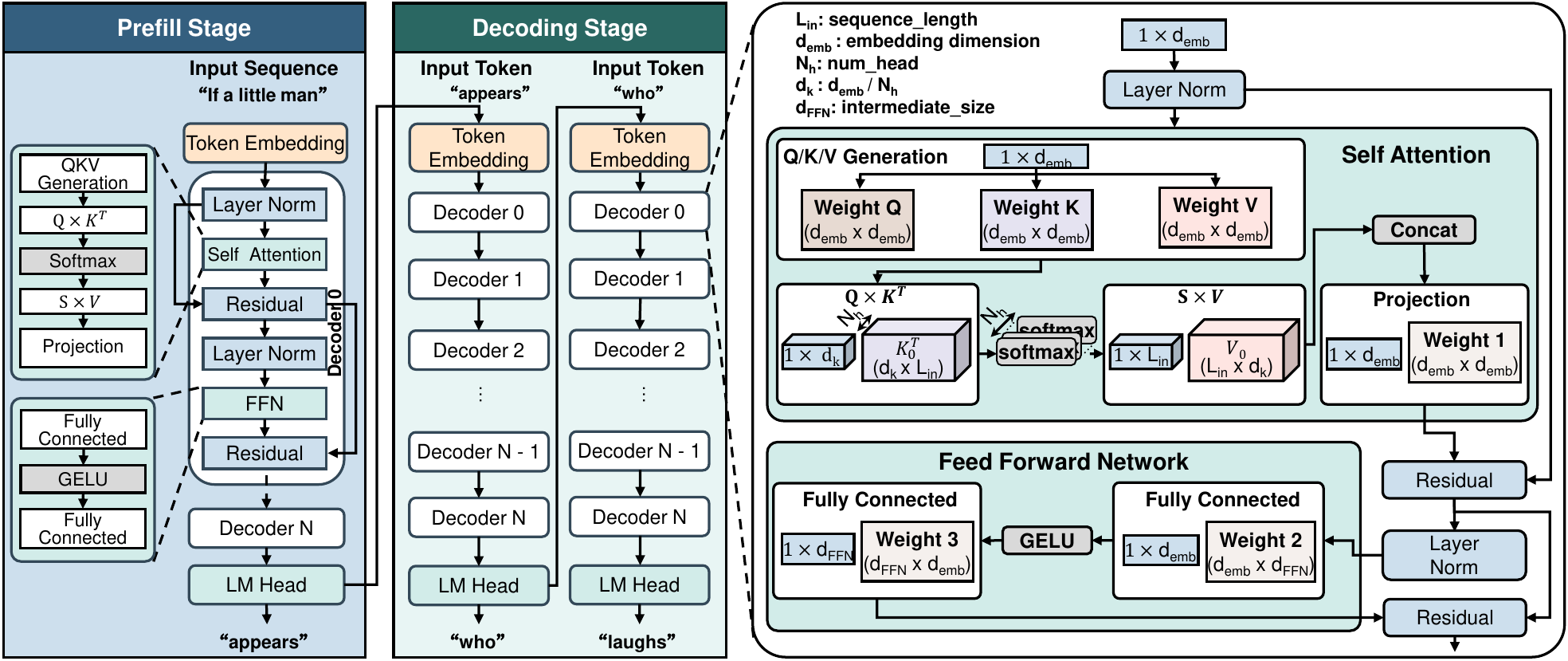}
\vspace{-5pt}
\caption{Model architecture and inference process of LLM.}
\label{fig2:Model Architecture}
\vspace{-12pt}
\end{figure*}
At the top of the pyramid, SRAM-PIM~\cite{qi2025ciminus, fu2023p,duan2025efficient} offers ultra-low latency ($\sim$$1 \times$), high internal bandwidth, and its compatibility with logic processes makes it particularly suitable for integrating complex functions.
However, due to limited storage density ($\sim$MB), SRAM-PIM alone is insufficient for meeting the massive memory requirements of LLM inference.
In contrast, HBM-PIM~\cite{he2025papi, li2024specpim} provides a balance of high bandwidth and capacity.
\hl{While PIM architectures reduce end-to-end access latency by eliminating data movement across the processor-memory interface, the effective latency remains constrained by intrinsic DRAM operations such as row activation and refresh cycles.
The resulting access latency typically resides in the range of tens of nanoseconds, approximately several tens of times that of SRAM-PIM.
This inherent latency gap, coupled with prohibitive logic integration costs, significantly hampers performance compared to SRAM-PIM solutions.
}
This fundamental integration challenge is also inherent to capacity-oriented technologies like LPDDR~\cite{park2024lpddr, kim2023samsung} and Flash~\cite{lee20223d, kang2021s}. 
Meanwhile, they are further hampered by substantially lower bandwidth and greater latency, rendering them unsuitable for high-performance LLM acceleration.

The above analysis reveals that a single memory technology cannot simultaneously satisfy the diverse requirements of LLM inference, including large model size, high bandwidth, computational diversity, and low latency.
This directly motivates the exploration of a heterogeneous PIM architecture to accelerate LLM inference by synergistically combining different memory technologies.
Based on our analysis, a combination of SRAM-PIM and HBM-PIM emerges as the most promising approach to meet these diverse demands.
Therefore, we propose HPIM, a heterogeneous PIM architecture, as depicted at the bottom of Fig.~\ref{fig1:PIM pyramid}.
This design leverages the complementary strengths of both memory types through a clear partitioning strategy.
This balanced approach offers a practical and scalable path forward for accelerating LLM inference.
Our contributions are as follows:

\begin{itemize}
\item 
\textbf{Memory-Centric Heterogeneous PIM Architecture:} 
We propose HPIM, the first memory-centric heterogeneous PIM architecture to integrate a low-latency, computationally flexible SRAM-PIM and a high-bandwidth, high-capacity HBM-PIM for single-batch LLM inference. 
This novel architecture leverages parallel execution between these two specialized subsystems to overcome the memory bandwidth bottlenecks and serial dependencies inherent in autoregressive decoding.

\item \textbf{Adaptive Workload Partitioning and Intra-Token Parallelism:} We introduce a novel hardware-aware workload partitioning and scheduling strategy that allocates compute-intensive, latency-sensitive tasks to SRAM-PIM and weight-intensive operations to HBM-PIM.
This strategy significantly minimizes inter-subsystem data movement and communication overhead.
Furthermore, to overcome token-level serialization, HPIM employs a tightly coupled execution pipeline across subsystems that exploits fine-grained intra-token parallelism.
This mechanism enables the overlapping of QKV generation, attention, and FFN operations, significantly shortening the decoding critical path.
\item 
\textbf{Comprehensive Evaluation:} 
\hl{We evaluate HPIM on a cycle-accurate simulator across model scales from the OPT family ($350$M$\sim$$30$B). 
Results show that HPIM achieves a peak speedup of up to $23.1\times$ compared to an NVIDIA A100 GPU.
Furthermore, HPIM achieves a speedup of up to 1.56× over IANUS and up to $4.88\times$ higher throughput than CXL-PNM, demonstrating its superiority over state-of-the-art (SOTA) PIM accelerators.}

\end{itemize}

\hl{The remainder of this paper is organized as follows. 
Section~\ref{sec: Background} reviews the background and the motivation for the HPIM accelerator. 
Section~\ref{sec: Overview} provides an overview of the HPIM.
Section~\ref{sec:5} and Section~\ref{sec: scheduling} illustrate the microarchitecture design of the HPIM and its strategies for workload mapping and execution scheduling, respectively. 
Section~\ref{sec: Experiments} presents a comprehensive evaluation of HPIM against state-of-the-art accelerators.
Section~\ref{sec:7} discusses related work and Section~\ref{sec: Conclusion} concludes the paper.}

\section{Background and Motivations}
\label{sec: Background}
\subsection{LLM Inference}\label{sec:2.1}

\begin{figure}[t]
    \centering
    \subfigure[Breakdown ($\%$).]{
        \centering
        \includegraphics[width=0.46\textwidth]{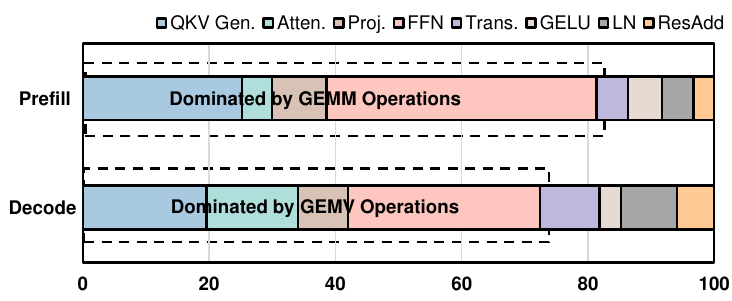}
        \label{Fig3-breakdown-a.pdf}
    }
    \subfigure[Breakdown ($\%$).]{
        \centering
        \includegraphics[width=0.47\textwidth]{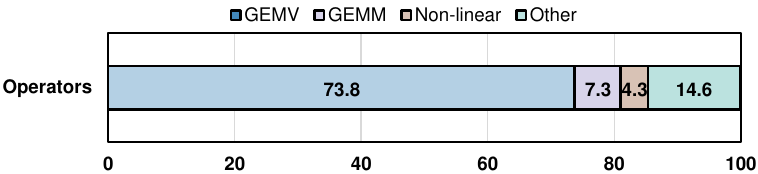}
        \label{Fig3-breakdown-b.pdf}
    }
    \vspace{-8pt}
    \caption{Execution breakdown of OPT-13B inference on an A100 GPU.
\hlblue{(a) Component-level breakdown across the prefill (GEMM-bound) and decode (GEMV-bound) stages.}
(b) Operator-level breakdown over the entire LLM inference process (e.g., GEMM, GEMV, Softmax).
The results are obtained using OPT-13B with an input length of 512 tokens and an output length of 32 tokens. It shows that the overall execution is overwhelmingly dominated by the GEMV-centric decode stage ($73.8\%$), highlighting it as the primary performance bottleneck.}
    \label{fig2:breakdown analysis at different strategy}
\end{figure}

Modern LLMs, predominantly based on the Transformer architecture~\cite{vaswani2017attention}, capture dependencies in text by computing relationships between all pairs of tokens in a sequence. 
\hl{As illustrated in Fig.~\ref{fig2:Model Architecture}, the execution of LLM inference on this architecture generally involves two distinct operational phases: prefill and decoding.
During the prefill phase, the model processes the entire input sequence in parallel, characterized by General Matrix-Matrix multiplication (GEMM), as shown in Fig.~\ref{Fig3-breakdown-a.pdf}.
In the subsequent decoding phase, the model generates tokens sequentially in an autoregressive manner, each prediction dependent on previously generated tokens.
The use of a Key/Value (KV) cache transforms the GEMM operations into a series of General Matrix-Vector (GEMV) operations.
This fundamental shift underscores the need for specialized hardware acceleration tailored specifically to these distinct computational patterns.}

\begin{figure}[t]
\centering
\includegraphics[width =0.7\linewidth]{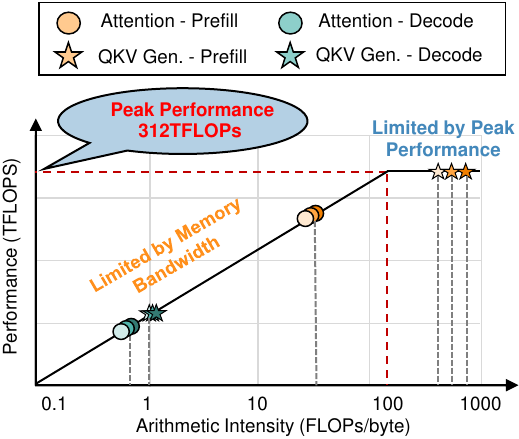}
\vspace{-8pt}
\caption{Roofline model of OPT-$6.7$B (Bright), OPT-$13$B (Moderate), and OPT-$30$B (Dark) operations on an A100 GPU with a sequence length of $2048$. The points plot the performance of Attention (circle) and QKV Generation (pentagram) during the prefill (orange) and decode (green) phases. }
\label{fig3:roofline_model}
\end{figure}

Traditional hardware accelerators, including GPUs, TPUs, and NPUs, have been extensively adopted for LLM inference due to their mature software ecosystem and optimized computational engines~\cite{rasley2020deepspeed, aminabadi2022deepspeed}. 
However, these architectures exhibit inefficiencies when executing single-batch LLM inference.
\hl{As depicted by the execution breakdown in Fig.~\ref{Fig3-breakdown-b.pdf}, memory-bound GEMV operations in the decoding stage account for 73.8\% of the total execution time, even in prefill-intensive scenarios with long input prompts.
The Roofline model in Fig.~\ref{fig3:roofline_model} elucidates that this decoding phase predominantly operates in the memory-bound regime, where performance is strictly constrained by memory bandwidth rather than peak computing capabilities. 
While conventional platforms are highly optimized for the GEMM-centric prefill phase, their performance degrades substantially when handling these bandwidth-sensitive GEMV kernels. 
Such a pronounced architectural mismatch identifies the decoding phase as the primary performance bottleneck, motivating the urgent necessity for the memory-centric PIM solutions proposed in this work.}
\subsection{Necessity of Heterogeneous PIM Design}
\textbf{\hl{Bottleneck Analysis.}}
\hl{Although the decoding phase is generally classified as memory-bound, a detailed analysis reveals that these underlying GEMV operations exhibit divergent behaviors in terms of memory access patterns and algorithmic complexity. Based on these intrinsic distinctions, we categorize the LLM inference workload into Weight-Intensive layers and Latency-Critical layers.
\begin{enumerate}
    \item \textbf{Weight-Intensive layers:} The Feed-Forward Networks (FFN) and Projection layers, often referred to as Fully Connected (FC) layers, constitute approximately 90\% of the model parameters.
    In single-batch inference, these layers perform GEMV operations using massive, static weight matrices accessed in a contiguous and sequential manner. 
    Furthermore, the computational complexity of these layers remains constant ($O(d^2)$) and purely linear, regardless of the sequence length $L$. 
    Consequently, we classify these layers as weight-intensive tasks, where the performance is strictly bounded by the memory bandwidth, rather than processing latency.
    \item \textbf{Latency-Critical Layers:}  In contrast, the Self-Attention mechanism exhibits compositional complexity, involving several heterogeneous operations including GEMV, matrix transpositions, and non-linear activations like Softmax.
    Unlike the static FC layers, the algorithmic complexity of Attention scales linearly with the sequence length ($O(L \times d)$), causing it to occupy an increasingly dominant portion of the critical path as the sequence expands.
    Additionally, retrieving KV pairs for these operations often involves irregular, non-contiguous memory accesses, which disrupt burst transfers and degrade effective bandwidth. Moreover, the Softmax operation necessitates a global reduction that prevents effective pipelining of subsequent operations. Due to these serial dependencies and dynamic complexities, the performance of the Attention mechanism is highly sensitive to latency. Thus, we define it as a latency-critical task.
\end{enumerate}}


\textbf{Serial Bottleneck in LLM Inference.}
As established in Sec.~\ref{sec:2.1}, single-batch LLM inference is dominated by the autoregressive decoding phase, where each token must be generated strictly after its predecessor.
Although prior PIM accelerators alleviate the bandwidth pressure of memory-bound GEMV kernels, they are inherently constrained by this \textbf{token-level serial dependency}.
To remove this barrier, we must pivot from inter-token to intra-token parallelism, the concurrent execution of sub-tasks within the generation of a single token.
This necessitates a composite heterogeneous PIM architecture where different types of PIM subsystems operate in parallel to accelerate these sub-tasks separately.
\hl{Specifically, the weight-intensive layers, characterized by their substantial static parameters, can be offloaded to a capacity-oriented unit (e.g., DRAM-PIM).
Conversely, the latency-critical attention layers are assigned to a speed-oriented one (e.g., SRAM-PIM).}
Such an architectural decomposition enables fine-grained pipelining and parallelism along the critical path of token generation, thereby directly addressing the serial execution bottleneck.

\textbf{Diversity in Computational Workloads.}
\hl{Beyond the dichotomy of weight-intensive and latency-critical tasks in the decoding phase, the end-to-end LLM inference execution encompasses an even wider spectrum of computational patterns.
First, the prefill phase relies on compute-intensive GEMM operations, where performance is bounded by arithmetic throughput rather than memory bandwidth.
Second, lightweight operations such as GELU, LayerNorm, and ResAdd introduce element-wise workloads.
These diverse operations exhibit varying demands in terms of memory bandwidth, storage capacity, and compute intensity.}
Consequently, adopting a homogeneous PIM architecture with a single memory technology proves insufficient. 
To mitigate this mismatch, a heterogeneous PIM architecture, integrating different memory technologies, is essential to concurrently satisfy the demands for high bandwidth, large storage capacity, and computational diversity, as shown in Fig.~\ref{fig1:PIM pyramid}. 
For example, SRAM-based PIM offers low-latency computation and high internal bandwidth but suffers from limited capacity; in contrast, DRAM- or Flash-based PIM provides large storage but incurs longer access latencies and consumes more energy.
Different tasks can be selectively assigned to the most suitable PIM domains based on their computational and memory characteristics. 
This strategic division improves resource utilization and load balancing.

 
\begin{figure*}[t]
\centering
\includegraphics[width =0.9\linewidth]{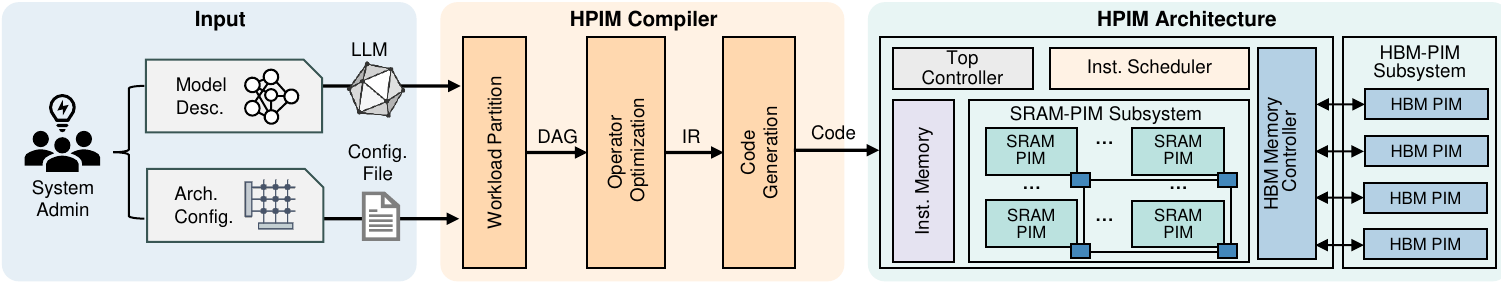}
\vspace{-5pt}
\caption{Overview of the HPIM accelerator, including its two main components: HPIM compiler and HPIM architecture. The workflow starts from user-defined model descriptions and architectural configurations, which are processed by the HPIM compiler to generate executable instructions. These instructions are then run on the HPIM architecture, which features a tightly coupled design of an SRAM-PIM subsystem and an HBM-PIM subsystem.}
\label{fig4:OverallArchitecturer}
\vspace{-12pt}
\end{figure*}
\subsection{Why HBM and SRAM?}\label{sec:2.5} Among various candidate memory technologies, we specifically select HBM and SRAM as the underlying media for our heterogeneous PIM-based accelerator, driven by the following key considerations:

\textbf{High Bandwidth Requirements.} The primary bottleneck of single-batch LLM inference lies in memory bandwidth constraints.
Among various memory technologies, HBM and SRAM inherently provide extremely high memory bandwidth. 
Moreover, integrating compute logic in memory banks or arrays enables even higher internal bandwidth utilization, substantially alleviating bandwidth-related performance bottlenecks.

\textbf{Cost Efficiency and Integration Feasibility.}
Despite its relatively recent emergence, SRAM-based PIM architectures have witnessed rapid development due to their excellent compatibility with standard digital processes.
This compatibility facilitates the seamless integration of full-core digital logic with SRAM-based PIM, enabling efficient implementation of complex operations, such as nonlinear activation functions. 
Conversely, embedding such intricate logic directly into DRAM or emerging non-volatile memory (NVM) technologies remains prohibitively challenging and costly. 
In such media, only simpler operations (e.g., multiply-accumulate, MAC) are realistically feasible. 
This delineation naturally positions SRAM as an optimal candidate for computationally intensive tasks requiring intricate logic integration.

\textbf{Reduction of Data Movement Overhead.} 
The weight matrices constitute a dominant fraction compared to the KV cache in single-batch inference scenarios. 
For instance, in an OPT-$30$B model with a $1k$ sequence length, the weights amount to approximately $60$ GB, while the KV cache is only about $1.31$ GB.
Extensive prior work demonstrates that frequent data movement is one of the primary sources of energy dissipation and performance loss.
Consequently, computations involving substantial weight matrices in QKV generation and FFN are strategically offloaded to HBM-based near-bank PIM to capitalize on the larger storage capacity and bandwidth advantages of HBM. 
In contrast, smaller-sized but computationally intensive operations, particularly attention computations and nonlinear operations, are effectively mapped onto SRAM-based PIM units, leveraging their unique characteristics of ultra-high speed, low latency, and high-throughput computation.

In summary, combining HBM-PIM with SRAM-PIM provides a highly practical, performance-efficient architecture for single-batch LLM inference.
HBM offers the bandwidth and capacity required for weight-intensive GEMV computations, while SRAM delivers ultra-low latency for attention and nonlinear operators. 
Our heterogeneous PIM design minimizes data movement, enables fine-grained pipelining and parallel execution, reduces end-to-end latency, and improves implementation feasibility.

\section{Overview of the HPIM Accelerator}\label{sec: Overview}
To address the critical bottlenecks in single-batch LLM inference, particularly the memory bandwidth limitations and serial dependencies in autoregressive decoding, we propose HPIM, a memory-centric heterogeneous PIM accelerator.
As depicted in Fig.~\ref{fig4:OverallArchitecturer}, HPIM comprises two tightly integrated components: a hardware-aware compiler framework and a heterogeneous hardware architecture.
\subsection{HPIM Compiler Framework}\label{sec:4.1}
The compiler framework bridges high-level LLM models with the underlying heterogeneous PIM architecture.
It takes the model graph and architectural configuration as inputs, performs a series of hardware-aware transformations, and generates optimized PIM instruction streams.
First, the compiler conducts operator analysis and annotation, tagging each node in the LLM graph based on its computational and memory characteristics (GEMV, GEMM, or nonlinear, etc.).
Recognizing distinct computational patterns, the compiler adopts a stage-specific operator mapping policy:
\begin{enumerate}
    \item \textit{Prefill Stage.} All computations, primarily large-scale GEMM kernels, are entirely dispatched to the SRAM-PIM subsystem. 
    This leverages the powerful processing capabilities of the SRAM-PIM subsystem, particularly its specialized compute units optimized for GEMM-intensive workloads.
    \item \textit{Decoding Stage.} The compiler strategically partitions the workload between the SRAM-PIM and HBM-PIM subsystems, enabling concurrent execution across these heterogeneous modules. 
    This balanced distribution effectively reduces pipeline stalls, minimizing serial dependencies. 
\end{enumerate}
Then, the compiler applies a hybrid tiling strategy, combining head-wise parallelism (HP) for multi-head attention and tensor-wise parallelism (TP) for spatially partitioning large matrices.
This strategy effectively balances computational workloads and minimizes inter-subsystem data movement. 
Subsequently, the optimized computation graph is translated into an intermediate representation (IR), capturing optimized operator mapping and execution schedules.
This IR is finally lowered into separate PIM-specific instruction streams for SRAM-PIM and HBM-PIM subsystems, including synchronization, data prefetching, and pipeline control instructions.
\subsection{HPIM Architecture}\label{sec:4.2}
The HPIM architecture comprises an SRAM-PIM subsystem, an The HBM-PIM subsystem, an HBM memory controller, a top controller, a hardware instruction (Inst.) scheduler, and on-chip Inst. memory. 
The SRAM-PIM subsystem is built upon a multi-core architecture. 
It comprises $32$ individual SRAM-PIM cores, facilitating high computational density and tightly-coupled high-bandwidth memory.
These cores, each with an independent instruction control flow, are interconnected by a high-throughput Network-on-Chip (NoC).
This organization facilitates scalable parallel execution across multiple attention heads or tensor matrices and enables flexible inter-core pipelining.
HBM-PIM subsystem consists of four HBM-PIMs, which function as high-capacity data storage and high-bandwidth parallel processors.
Each HBM-PIM stores large model weights and the KV cache required for generative inference.
Lightweight arithmetic circuits are embedded at the DRAM-bank level so that multiplications and accumulations are performed close to the data while reducing area and power overhead.
Meanwhile, to facilitate seamless data flow between the two subsystems, each SRAM-PIM core is equipped with a dedicated interface that provides direct access to specific HBM channels, enabling efficient, direct data transfers.
Furthermore, a centralized top-level controller combined with a hardware-based instruction scheduler dynamically orchestrates task execution, optimally pipelines computations across SRAM and HBM units, and efficiently synchronizes data movement. 

In summary, by intelligently partitioning tasks based on hardware advantages, employing a sophisticated compiler strategy, and leveraging deep pipelining between heterogeneous subsystems, HPIM significantly reduces inference latency and maximizes throughput in single-batch LLM inference workloads.
\begin{figure}[t]
\centering
\includegraphics[width =0.8\linewidth]{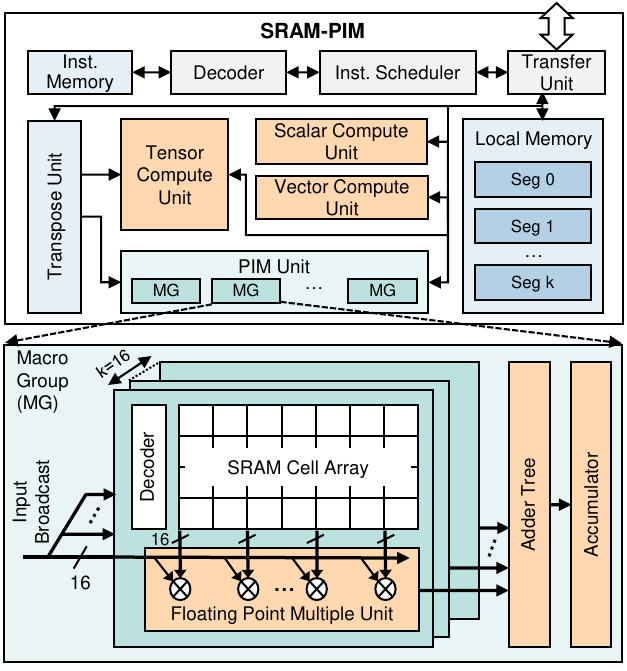}
\vspace{-6pt}
\caption{The microarchitecture of the SRAM-PIM subsystem, which is designed as a computing engine (Tensor, Vector, Scalar, and PIM Units) within the HPIM architecture.}
\label{fig6:SRAM-PIM architecture}
\end{figure}

\begin{figure*}[t]
\centering
\includegraphics[width =0.9\linewidth]{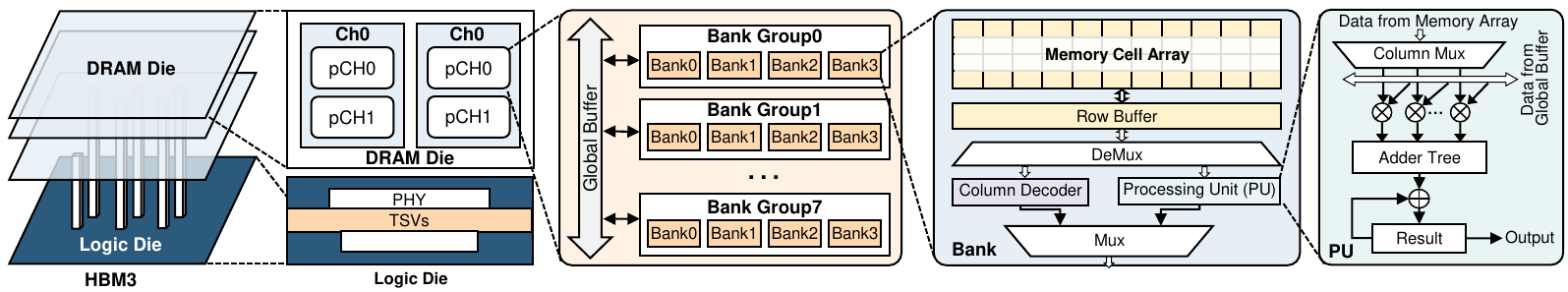}
\vspace{-8pt}
\caption{\hlblue{The microarchitecture of the HBM-PIM subsystem. It illustrates the hierarchical structure based on HBM3, from the 3D-stacked package down to the in-bank Processing Unit (PU).}}
\label{fig5:HBM-PIM Architecture}
\vspace{-14pt}
\end{figure*}
\section{Microarchitecture Design of HPIM}\label{sec:5}
\hblue{This section elaborates on the microarchitectural details of the two subsystems outlined in Section~\ref{sec:4.2}.}
\subsection{SRAM-Based PIM Design}\label{sec:5.1}
As shown in Fig.~\ref{fig6:SRAM-PIM architecture}, the SRAM-PIM subsystem functions as a high-performance computing engine composed of computational units, storage modules, and a sophisticated instruction scheduling module.

\textbf{Computational Module.} Each SRAM-PIM core incorporates multiple specialized computational units to meet the different computation requirements in LLM inference.
To achieve high internal bandwidth, the PIM unit performs computation directly within SRAM arrays. 
Each unit consists of $16$ macro groups (MGs) that share the same inputs and are equipped with integrated floating-point multipliers, adder trees, and accumulators.
However, the PIM unit solely performing GEMV operations is insufficient for complete LLM inference. 
To handle the compute-intensive GEMM operations prevalent in the prefill stage, each core includes a powerful tensor compute unit (TCU), which is realized by a $64 \times 64$ systolic array for high-throughput matrix multiplication.
Concurrently, a programmable vector compute unit (VCU) is responsible for essential element-wise operations. 
It supports a range of scalar and vector-level operations, including addition, subtraction, multiplication, division, exponentiation, and square root. 
These capabilities are crucial for implementing complex non-linear functions like Softmax and GELU, normalization layers like LayerNorm, and residual additions.
The scalar compute unit (SCU) focuses on executing scalar arithmetic and managing control flow operations.

\textbf{Instruction Scheduling Module.} 
To support diverse workloads and simplify compiler design, the SRAM-PIM core implements a unified $32$-bit instruction set architecture (ISA) with specialized variants for different operation types.
Instructions are categorized into three main classes: compute, communication, and control flow instructions.
The compute instructions are further specialized for the PIM unit, TCU, VCU, and SCU. 
The execution of this instruction set is managed by a three-stage pipeline (Fetch-Decode-Execute) designed to maximize throughput. 
The process begins at the Fetch stage, where compiler-generated instructions are retrieved from instruction memory. During the subsequent Decode stage, each instruction is parsed and dispatched to a dedicated instruction scheduler.
This scheduler is central to the Execute stage, where it dynamically analyzes data dependencies, monitors the status of the various compute units, and orchestrates task execution to maintain high pipeline utilization.

\textbf{Storage Module.} The storage module encompasses local memory for input/output data caching, a dedicated transpose unit, and a transfer unit. 
The transpose unit efficiently transforms the $K$ matrix retrieved from DRAM into an appropriate format for subsequent computation within the PIM unit. 
The transfer unit facilitates efficient data movement between various computational and storage modules within the SRAM-PIM subsystem, ensuring streamlined communication and data accessibility. 

Overall, these tightly integrated components substantially improve the efficiency, scalability, and overall performance of the SRAM-PIM subsystem within the HPIM architecture.

\subsection{HBM-Based PIM Design}\label{sec:3.2.2}
The HBM-PIM subsystem, illustrated in Fig.~\ref{fig5:HBM-PIM Architecture}, leverages an advanced HBM3 stack architecture comprising $8$ vertically stacked 8-Hi DRAM dies interconnected by TSVs.
Each DRAM die contains $2$ independent channels, each with $2$ pseudo-channels (pCH).
These pCHs are further subdivided into $8$ bank groups (BG) and $4$ banks, facilitating massive internal bandwidth and highly parallel data access.
To avoid the prohibitive cost of integrating complex, full-core logic into HBM, our HBM-PIM subsystem adopts simplified compute logic integrated at the bank level.
Building upon previous PIM paradigms (e.g., Newton architecture~\cite{he2020newton}), each bank integrates dedicated multiply-accumulate (MAC) units to process large-scale, weight-intensive GEMV computations directly near the memory cell arrays, significantly reducing data movement overhead.
In contrast to the SRAM-PIM subsystem, the HBM-PIM units are managed by a set of simple commands rather than a full-fledged ISA.
This design choice minimizes control overhead and hardware complexity within the bank.
\vspace{-2pt}

\hblue{The proposed DRAM die supports two classes of operations: conventional memory access (i.e., read and write operations) and PIM-enabled computation.
Both types of operations share a unified memory controller and command path, with the distinction occurring at the command interpretation and scheduling levels.
In the PIM execution path, input activations are initially loaded into the global buffer shared by all banks within a pCH, and then simultaneously broadcast to each bank.
Concurrently, weight and activation streams are processed in parallel by PUs, efficiently performing multiplication and accumulation operations.
To further minimize unnecessary data movement, our design fully exploits input reuse by ensuring input elements across matrix rows are loaded only once. 
The partial sums resulting from computations are temporarily stored within dedicated result units, which subsequently either transfer the final results externally or write them back into HBM storage. 
Furthermore, PIM-aware scheduling ensures execution without violating intrinsic DRAM refresh timing constraints.
Overall, by improving data locality, reducing redundant data movement, and exploiting bank-level parallelism, the proposed design significantly enhances the computational efficiency of the HBM-PIM subsystem.} 

\begin{figure*}[th]
\centering
\includegraphics[width =0.9\linewidth]{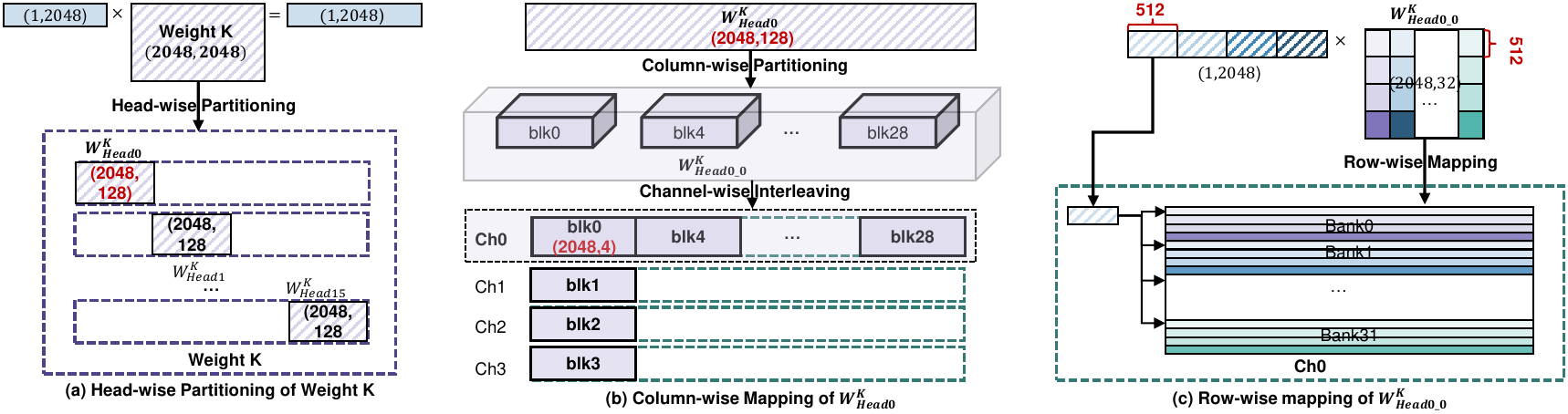}
\vspace{-5pt}
\caption{Illustration of the allocation and tiling scheme for the K weights of 16 heads in the HBM-PIM subsystem.}
\label{Fig8-mapping-strategy-for-HBM-PIM}
\vspace{-8pt}
\end{figure*}

In summary, the proposed heterogeneous architecture of HPIM efficiently partitions computation across SRAM-PIM and HBM-PIM subsystems based on latency sensitivity and bandwidth demands. SRAM-PIM emphasizes low-latency, versatile computation for GEMM, GEMV, and nonlinear operations, while HBM-PIM focuses on maximizing internal memory bandwidth for weight-intensive GEMV operations. 
Coordinated by a centralized instruction scheduler and optimized by the compiler, this synergistic design significantly enhances performance, efficiency, and scalability for single-batch LLM inference.

\section{Hardware-Aware Scheduling of HPIM}\label{sec: scheduling}
This section focuses on the hardware-aware scheduling of HPIM systems. 
We first elaborate on the workload mapping techniques of the weight matrices and attention layers between the heterogeneous HBM-PIM and SRAM-PIM subsystems. 
Then, we detail the scheduling strategy derived from these mappings, focusing on pipeline design tailored for multi-head self-attention, load balancing considerations, and coordinated optimization across prefetching, computation, and data transmission tasks.
\subsection{Workload Mapping}\label{sec:6.1}
Modern LLMs exhibit significant variability in the number of attention heads, typically ranging from $16$ to $128$.
To maximize resource utilization and improve the scalability of the various LLM models, we adopt a hybrid parallelism strategy that combines HP and TP across both the HBM-PIM and SRAM-PIM subsystems.

\begin{algorithm}[th]
\caption{Hybrid Parallelism of Q/K/V Weight Allocation}
\label{alg:head_tensor_partition}
\begin{algorithmic}[1]
\Require
Number of heads \( N_h \), DRAM channels \( N_D \), SRAM cores \( N_S \), weight dimensions \( d_{emb} \), tensors  \( \{T_{h}\} \)  for each head \( h \).

\Ensure
Partitioned tensor slices \( \{S_{c, r}\} \) to each channel \( c \) and round \( r \).
\State{\textcolor{brown}{// Calculate dimensions per head}}
\State \( d_k \gets d_{emb} / N_h \) 
\State {\textcolor{brown}{// Initialize counters for allocated heads and rounds}}
\State \( h_{\text{idx}} \gets 0 \) 
\State \( r \gets 0 \)

\While{ \( h_{\text{idx}} < N_h \) }
    \State{\textcolor{brown}{// Calculate the number of remaining heads}}
    \State \( h_{\text{rem}} \gets N_h - h_{\text{idx}} \) 
    \If{ \( h_{\text{rem}} <= 0 \) } \textbf{break} \EndIf
    \State{\textcolor{brown}{// Calculate the number of heads for HP in this round}}
    \State \( h_r \gets \min(h_{\text{rem}}, N_D, N_S) \)
    \State \( h_p \gets 2^{\lfloor \log_2 h_r \rfloor} \) 
    \State{\textcolor{brown}{// Calculate allocated channels per head for TP}}
    \State \( N_{ch} \gets N_D / h_p \)

    \For{ \( h = h_{\text{idx}} \) to \( h_{\text{idx}} + h_p - 1 \) }
        \State{\textcolor{brown}{// Apply channel-wise interleaving}}
        \For{ \( i = 0 \) to \(  d_k \) } 
            \State \( c \gets (h-h_{idx})\cdot N_{ch} + (i \% N_{ch}) \) 
            \State \( S_{c, r} \gets \text{Append}(T_h[i]) \)
        \EndFor
    \EndFor
    
    \State \( h_{\text{idx}} \gets h_{\text{idx}} + h_p \)
    \State \( r \gets r + 1 \)
\EndWhile
\end{algorithmic}
\end{algorithm}

\hl{On the HBM-PIM side, the weight matrices of Q, K, and V employ this hybrid parallelism strategy, as depicted in Alg.~\ref{alg:head_tensor_partition} and Fig.~\ref{Fig8-mapping-strategy-for-HBM-PIM}.
The detailed mapping strategy is as follows:
\begin{enumerate}
    \item \textbf{Partitioning via HP:} The process initiates by partitioning large weight matrices (Q/K/V) along the head dimension to exploit coarse-grained parallelism. As depicted in Fig.~\ref{Fig8-mapping-strategy-for-HBM-PIM}(a), a large weight matrix (e.g., $W^K$) is decomposed into independent sub-matrices (e.g., $W_{Head0}^K$), each associated with a distinct attention head. To ensure adaptability under varying resource constraints, the system dynamically determines the optimal degree of parallelism, denoted as $h_p$, as derived in Alg.~\ref{alg:head_tensor_partition}.
    \item \textbf{Allocation via TP:} Subsequently, these partitions are mapped to physical HBM channels using TP to maximize bandwidth utilization. We define $N_{ch}$ as the number of channels allocated per head.
    To ensure load balancing, we employ a column-wise tiling and interleaving strategy.
    Specifically, weight matrix columns are distributed across channels using modulo arithmetic ($c \leftarrow \dots + (i \% N_{ch})$).
    This distribution results in the interleaved pattern depicted in Fig.~\ref{Fig8-mapping-strategy-for-HBM-PIM}(b). In the illustrated instance where $N_{ch}=4$, consecutive weight blocks (\emph{blk0} through \emph{blk3}) are striped across channels \emph{Ch0} through \emph{Ch3}. This interleaving strategy ensures that sequential data access activates multiple channels concurrently, thereby saturating the aggregate internal bandwidth.
    \item \textbf{Bank-level Allocation:} Finally, within each allocated channel, data is mapped to specific DRAM banks to optimize local access latency. As illustrated in Fig.~\ref{Fig8-mapping-strategy-for-HBM-PIM}(c), we adopt a row-wise mapping scheme aligned with the DRAM row-buffer size. This layout minimizes row-buffer conflicts and ensures that the data access granularity matches the requirements of the near-bank processing units.
\end{enumerate}
For weights associated with other FC layers (e.g., \texttt{weight1}, \texttt{weight2}, and \texttt{weight3} in Fig.~\ref{fig2:Model Architecture}), we leverage TP by slicing large weight matrices along spatial or row dimensions.
These matrices are then mapped across multiple HBM channels, significantly reducing data movement, balancing computational loads, and matching the access granularity of DRAM banks.}

\begin{figure*}[htbp]
\centering
\includegraphics[width =0.8\linewidth]{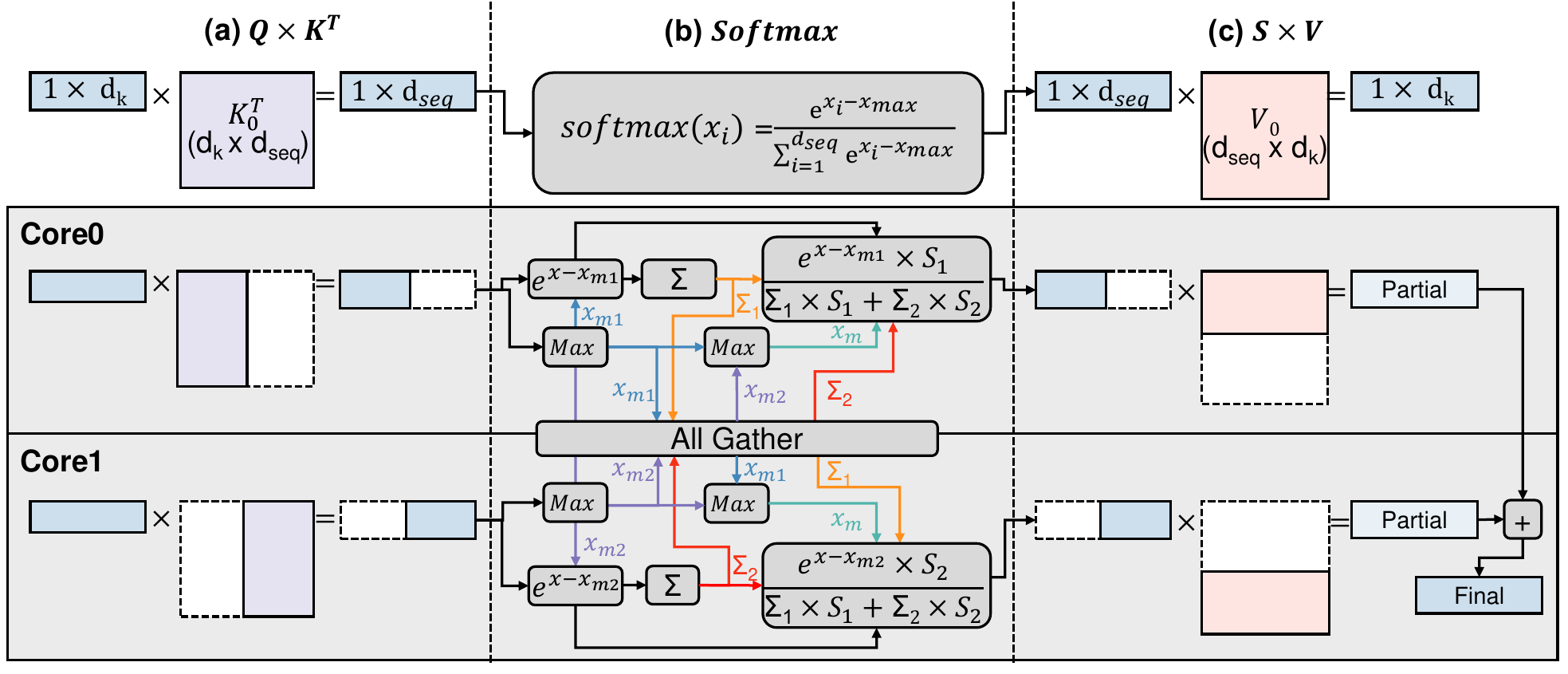}
\vspace{-5pt}
\caption{Illustration of parallel processing of a single head on multiple SRAM-PIM cores. The figure illustrates the three main stages of attention: (a) GEMV operation between the query vector and the transposed key matrix ($Q \times K^T$), (b) softmax computation with distributed max and sum operations across cores and subsequent All-Gather synchronization, and (c) GEMV operation of the attention scores with the value matrix ($S \times V$).}
\label{Fig9-Mapping-SRAM}
\vspace{-12pt}
\end{figure*}
On the SRAM-PIM side, we adopt a combination of HP and intra-head TP to accommodate varying model configurations efficiently. 
Each head can be executed either independently on a single core or collaboratively across multiple cores to maximize parallelism and resource utilization.
For example, when the number of heads matches the number of available compute cores (e.g., 32 heads on 32 cores), each head is exclusively mapped to a single core.
This arrangement allows entirely localized attention computation, including softmax normalization, eliminating inter-core communication and minimizing latency.
For larger models with more heads than available cores, computations are partitioned and scheduled across multiple execution phases.

Conversely, when the number of heads is fewer than the available compute cores, we apply intra-head TP by evenly distributing the computation of a single head across multiple cores.
As illustrated in Fig.~\ref{Fig9-Mapping-SRAM}, each core handles a distinct tensor partition and computes partial results independently.
Then, we apply an efficient All-Gather to recover the global normalization factors for softmax. 
This strategy preserves computation locality while maintaining correctness with minimal NoC overhead.

\subsection{Pipelined Parallelism for LLM Inference}\label{sec:5.2}

\begin{figure}[t]
    \centering
    \subfigure[Prefill phase.]{
        \centering
        \includegraphics[width=0.48\textwidth]{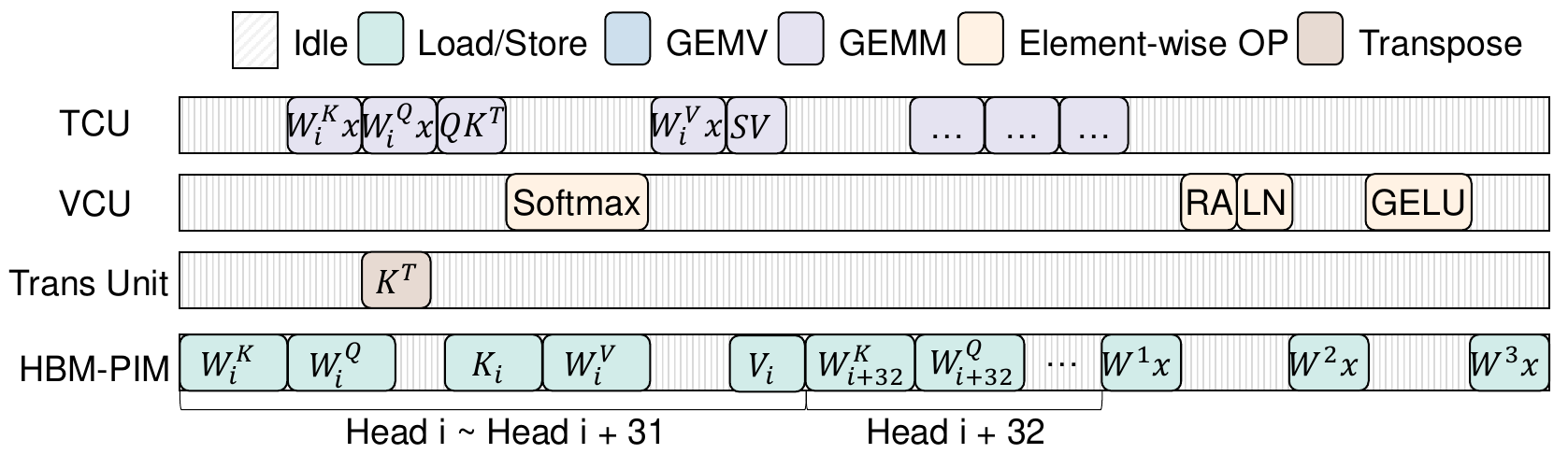}
        \label{fig:Prefill phrase}
    }
    \subfigure[Decoding phase.]{
        \centering
        \includegraphics[width=0.48\textwidth]{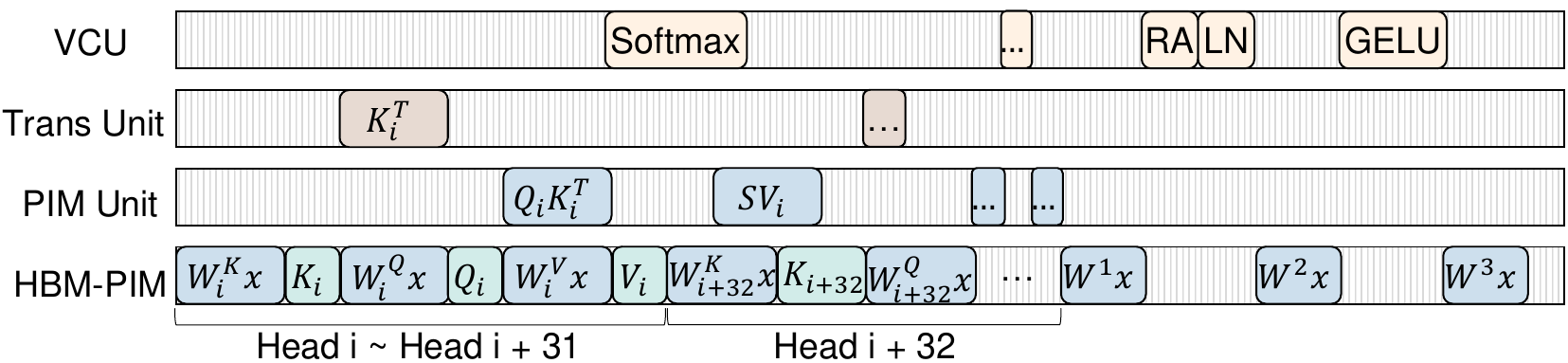}
        \label{fig: Decoding phrase}
    }
    \vspace{-10pt}
    \caption{Example execution timelines for a 32-head model in (a) the prefill phase, where FCs are mapped to the TCU, and (b) the decoding phase, where FCs are mapped to the PIM unit. TCU: tensor compute unit, VCU: vector compute unit, Trans Unit: transpose unit, PIM Unit: PIM unit in SRAM-PIM core.}
    \label{fig11:breakdown analysis at different strategy}
\end{figure}

\hl{Given that the workload of the prefill phase is dominated by large-scale GEMM kernels, HPIM dispatches these compute-intensive operations to the TCU in the SRAM-PIM core to exploit its superior arithmetic throughput.
Concurrently, element-wise non-linear kernels such as layer normalization, residual addition, GELU, and the softmax are offloaded to the VCU.
To hide memory latency, the architecture utilizes a tightly coupled execution pipeline shown in Fig.~\ref{fig:Prefill phrase}.
Specifically, weight loading from HBM-PIM is overlapped with computation, while the $Q$ matrix generation runs in parallel with the $K$ matrix transposition. 
This enables the rapid initiation of attention scores ($Q \times K^T$), which are immediately pipelined to the VCU for Softmax normalization while $W^V$ weights are prefetched.}

During the decoding phase, the main workload shifts toward memory-bound GEMV operations, and memory bandwidth becomes the primary bottleneck. 
To alleviate the bandwidth limitation, weight-intensive FC layers are migrated to the HBM near-bank PIM subsystem, while latency-critical attention kernels and non-linear kernels remain in the SRAM-PIM subsystem. 
Since $Q \times K^T$ and $S \times V$ are also GEMV operations, we deliberately execute them in the PIM Unit of the SRAM-PIM subsystem to enable the computation pipeline.
As shown in Fig.~\ref{fig: Decoding phrase}, the decoding execution can be broken down as follows:
\begin{itemize}
    \item \textbf{Bandwidth-Critical GEMVs in HBM-PIM.}
    All weight-intensive GEMV computations, including $Q/K/V$ generation and FFN layers, are executed within near-bank HBM-PIM units.
    Each channel computes the $K$, $Q$, and $V$ vectors for its assigned heads sequentially and streams them to the corresponding SRAM-PIM cores.
    \item \textbf{Pipelined Attention.}
    To minimize idle time during execution, SRAM-PIM subsystems are tightly pipelined with HBM-PIM execution.
    For example, once the $K$ vector is generated and streamed into the SRAM-PIM core, the transpose unit immediately begins converting $K$ into $K^T$ in parallel with the ongoing generation of $Q$ in HBM-PIM.
    Likewise, the computation of $Q \times K^T$ in PIM Unit is overlapped with the generation of $V$.
    This tightly coupled pipelining between SRAM-PIM and HBM-PIM execution shortens the critical path and improves throughput.
    If multiple cores are assigned to a single head, we perform an all-gather softmax reduction, where only local maxima and exponent sums are exchanged to minimize NoC traffic.
    The resulting attention scores are forwarded back to the PIM units for the $S \times V$ GEMV, while the HBM-PIM fabric concurrently computes and transforms the next head group.
    \item \textbf{Projection and FFN Distribution.}
    After attention, the subsequent projection and FFN layers, which involve weight-heavy GEMV operations, are still offloaded to the near-bank HBM-PIM subsystem.
    The corresponding weight matrices are pre-partitioned and striped across HBM channels to enable parallel execution. 
    This channel-aligned mapping ensures load balancing across memory stacks.
\end{itemize}

\section{Evaluation Results}\label{sec: Experiments}

\subsection{Experimental Setup}
\textbf{LLM Models.} We evaluate a series of transformer-based LLMs from the OPT family~\cite{zhang2022optopenpretrainedtransformer}, with configurations ranging from $350$M to $30$B parameters, as detailed in Tab.~\ref{tab2}. 
This allows us to explore the scalability and efficiency of HPIM across diverse LLM configurations.
All baseline models are executed using FP16 precision under a single-batch (batch size = 1) inference setting.
While our evaluation focuses on the OPT family as the benchmark, our proposed HPIM architecture is model-agnostic and can be readily extended to other transformer-based LLMs.
All baseline workloads are executed using the Hugging Face Transformers framework~\cite{wolf2019huggingface}.

\textbf{Baseline.} 
We evaluate HPIM against a set of state-of-the-art baselines, including the NVIDIA A100 GPU, and two representative PIM-related accelerators: the IANUS hybrid NPU-PIM system~\cite{seo2024ianus} and the CXL-PNM near-memory~\cite{park2024lpddr} platform.
\hl{The detailed specifications of each architecture are summarized in Tab.~\ref{tab4}.
Notably, the table reports the aggregated internal bandwidth for PIM-based systems, a critical metric inherent to the PIM paradigm.
The internal bandwidth is calculated as:
\begin{equation}
    Internal \ Bandwidth=N_{ch} \times N_{bank} \times BCIW \times f,
\end{equation}
where $N_{ch}$ denotes the number of channels, $N_{bank}$ represents the number of banks per channel, $BCIW$ indicates the bit width of the internal computing unit per bank, and $f$ is the operating frequency.}

\begin{table}[tp]
\centering
\caption{LLM configurable parameters.}
\label{tab2}
\resizebox{\linewidth}{!}{
\begin{tabular}{|cc|c|c|c|c|c|}
\hline
\multicolumn{2}{|c|}{Models}                       & \begin{tabular}[c]{@{}c@{}}Embedding\\ dimension\end{tabular} & Layers & Heads & Params & d\_k \\ \hline
\multicolumn{1}{|c|}{\multirow{5}{*}{OPT}} & 350M & 1024                                                          & 24     & 16   & 350M   & 64           \\ \cline{2-7} 
\multicolumn{1}{|c|}{}                     & 1.3B  & 2048                                                          & 24     & 32    & 1.3B   & 64           \\ \cline{2-7} 
\multicolumn{1}{|c|}{}                     & 6.7B  & 4096                                                          & 32     & 32    & 6.7B   & 128          \\ \cline{2-7} 
\multicolumn{1}{|c|}{}                     & 13B   & 5120                                                          & 40     & 40    & 13B    & 128          \\ \cline{2-7} 
\multicolumn{1}{|c|}{}                     & 30B   & 7168                                                          & 48     & 56    & 30B    & 128          \\ \hline
\end{tabular}
}
\vspace{-8pt}
\end{table}

\begin{table}[tp]
\centering
\caption{\hl{Specifications of A100, IANUS, CXL-PNM, HPIM, and HPIM-S.}}
\label{tab4}
\begin{threeparttable}
\resizebox{\linewidth}{!}{
\begin{tabular}{|c|c|c|c|c|c|}
\hline
                                                                                   & A100  & IANUS & CXL-PNM & HPIM  & \hl{HPIM-S} \\ \hline
Frequency (MHz)                                                                    & 1155  & 700   & 1024    & 800  & \hl{800}   \\ \hline
\begin{tabular}[c]{@{}c@{}}\hl{Peak Tensor}\\ \hl{Engine Throughput}\\ \hl{(TFLOPS)}\end{tabular} & 321   & 184   & 4.09    & \hl{209.7$^\ast$}   & \hl{209.7}    \\ \hline
\begin{tabular}[c]{@{}c@{}}SRAM Capacity\\ (MB)\end{tabular}                       & 84    & 64    & 64      & 45    & \hl{45}     \\ \hline
DRAM Type                                                                          & HBM2e & GDDR6 & LPDDR   & HBM3  & \hl{HBM3}   \\ \hline
\begin{tabular}[c]{@{}c@{}}DRAM Capacity\\ (GB)\end{tabular}                       & 80    & 8     & 512     & 96    & \hl{48}     \\ \hline
\begin{tabular}[c]{@{}c@{}}DRAM Bandwidth \\ (GB/s)\end{tabular}                   & 1935  & 256   & 1100    & 3276.8  & \hl{1638.4}   \\ \hline
\begin{tabular}[c]{@{}c@{}}DRAM Internal \\ BandWidth (TB/s)\end{tabular}          & N/A   & 4.09     & N/A     & \hl{26.2} & \hl{13.1}   \\ \hline
\end{tabular}
}
\begin{tablenotes}[flushleft]
    \footnotesize
    \item \hl{$\ast$ The 209.7 TFLOPS refers to the specialized TCU in SRAM-PIM.}
\end{tablenotes}
\end{threeparttable}
\vspace{-6pt}
\end{table}

\begin{table}[tp]
\centering
\caption{\hl{Hardware configuration of HPIM.}}
\label{tab3}
\resizebox{1.0\linewidth}{!}{
\begin{tabular}{|c|c|c|}
\hline
    HPIM                  & Composition          & \begin{tabular}[c]{@{}c@{}}4 HBM-PIM\\ 32 SRAM-PIM, 800 MHz Frequency\end{tabular} \\ \hline
\multirow{6}{*}{SRAM-PIM} & TCU          &   64*64 PEs, 1 MAC per PE \\ \cline{2-3} 
                          & VCU          &  \begin{tabular}[c]{@{}c@{}}   Add, Sub, Mult, Div,\\  Exp, Square Root, etc.  \end{tabular} \\ \cline{2-3} 
                          & PIM Unit           & \begin{tabular}[c]{@{}c@{}}  16 MG, 16 SRAM-PIM Macro per MG, \\ 4KB per SRAM-PIM Macro, \\ 8 FP16 Multipliers per Macro, \hl{3.28} TFLOPS                                                 \end{tabular}                            \\ \cline{2-3} 
                          & Local Memory                & \begin{tabular}[c]{@{}c@{}} 384KB Activation Memory \\  32KB Temp Memory \end{tabular}                 \\ \hline
\multirow{5}{*}{HBM-PIM}  & HBM3 Configuration    & \begin{tabular}[c]{@{}c@{}}8 dies/HBM, 24Gb/die, 8-Hi DRAM, \\ 2 Channels per DRAM, \\ 8 Bank Groups (BGs) per Channel, \\ 4 Banks per BG, 1KB Page Size per pCH \\
\hl{Data transfer rate is 6.4Gbps}\end{tabular}  \\ \cline{2-3} 
                          & Processing Unit (PU) & \begin{tabular}[c]{@{}c@{}} \hl{400MHz, 1 PU per Bank, }\\ \hl{8 FP16 Multipliers per PU}  \\
                          \end{tabular}      \\ \cline{2-3} 
                          & Global Buffer        & One 1 KB Global Buffer per pCH        \\ \cline{2-3} 
                          & \hl{Extra Area} & \hlblue{3.88 $mm^2$}  \\ \cline{2-3} 
                          \hline
\end{tabular}
}
\end{table}

\begin{figure*}[t]
\centering
\includegraphics[width =0.8\linewidth]{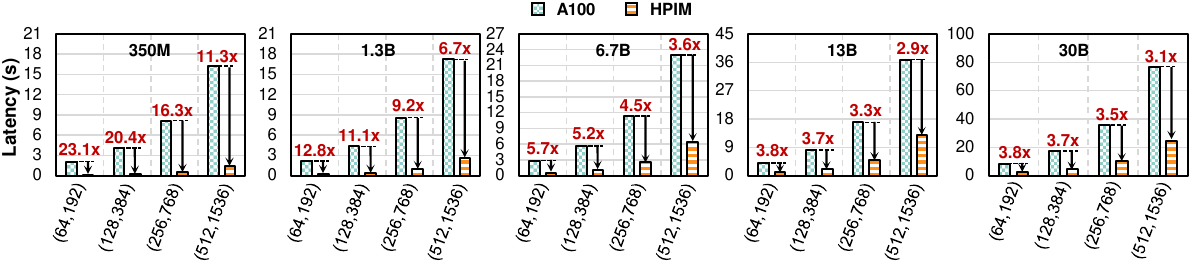}
\vspace{-8pt}
\caption{\hl{End-to-end inference latency comparison between HPIM and an NVIDIA A100 GPU across a range of OPT models ($350$M to $30$B) and sequence lengths.}}
\label{Fig11-latency-breakdown}
\vspace{-18pt}
\end{figure*}

\textbf{Hardware Specification of HPIM Architecture.} 
\hl{The hardware configuration of our proposed HPIM architecture is detailed in Tab.~\ref{tab3}. 
The system integrates $32$ SRAM-PIM cores with $4$ HBM-PIM modules, interconnected via a standard 2.5D silicon interposer using the HBM3 physical interface (PHY). Each SRAM-PIM core is a powerful processing engine containing several specialized compute units (TCU, VCU, SCU, PIM Unit) and local memory, operating at $800$~MHz. 
The HBM-PIM subsystem is built upon an HBM3-based PIM design.
To support large-scale LLM inference (e.g., 30B parameters), the system incorporates four HBM3 stacks, providing a total aggregate capacity of $96$~GB.
Each HBM3 module is configured as an 8-Hi stack using $24$~Gb dies with $16$ channels per stack. 
With a data rate of $6.4$~Gbps per pin across $1024$ data lanes per stack, the system achieves a peak aggregate bandwidth of $3.2$~TB/s. 
Considering the thermal constraints of 3D-stacked memory, the Processing Unit (PU) within the HBM-PIM subsystem is configured to operate at $400$~MHz.}

\textbf{HPIM Simulator.} We evaluate the proposed HPIM architecture using a cycle-accurate simulator capable of capturing detailed compute and memory behaviors across both the SRAM-PIM and HBM-PIM subsystems.
\hl{For the SRAM-PIM subsystem, we developed a custom cycle-accurate simulator. The digital circuits in the SRAM-PIM subsystem are implemented using Verilog HDL. 
The area and power are derived from synthesis results using Synopsys Design Compiler at a 12nm technology.}
The HBM-PIM subsystem is modeled using an extended version of the open-source DRAMsim3 simulator~\cite{li2020dramsim3}.
This extended simulator supports bank-level compute execution and PIM-aware scheduling, with timing and power parameters configured for HBM3 memory~\cite{HBM3}.
\hl{To ensure accurate hardware overhead and timing analysis, we synthesized the Processing Unit (PU) and Global Buffer using the same 12nm technology. The area results of these units in HBM-PIM are presented in Tab.~\ref{tab3}. }
\hl{We build a simulator based
on an open-source NoC-simulator Noxim~\cite{catania2016energy} to evaluate the performance of NoC.}


\subsection{Performance Evaluation}\label{sec: Evaluation of FTA Algorithm}
\textbf{End-to-End Inference Latency.} \hl{We first evaluate the end-to-end inference latency and speedup of our proposed HPIM architecture compared to an NVIDIA A100 GPU.
As illustrated in Fig.~\ref{Fig11-latency-breakdown}, HPIM consistently achieves significantly lower inference latency across all tested model scales and sequence lengths, delivering a peak speedup of up to $23.1\times$.
For the LLM inference workload with significantly more output tokens than input tokens, i.e., (256,768), HPIM demonstrates $4.5\times$, $3.3\times$, and $3.5\times$ lower latency than the A100 GPU for the OPT-6.7B, OPT-13B, and OPT-30B models, respectively.
This performance gain is particularly significant considering that the peak compute throughput of the PIM module ($26.2$ TFLOPS for HBM-PIM and $104.9$ TFLOPS for PIM unit in SRAM-PIM) is substantially lower than that of the A100 GPU ($312$ TFLOPS).}
This result underscores a fundamental insight: single-batch inference performance is dictated by architectural efficiency in the memory-bound decoding stage, not just peak compute throughput.
By unlocking intra-token parallelism, HPIM concurrently executes different GEMV tasks across its SRAM-PIM and HBM-PIM subsystems. This memory-centric, parallel approach directly mitigates the data movement and serial dependency bottlenecks that constrain conventional accelerators, demonstrating the superiority of its design for latency-critical applications.

\begin{figure}[t]
\centering
\includegraphics[width =0.8\linewidth]{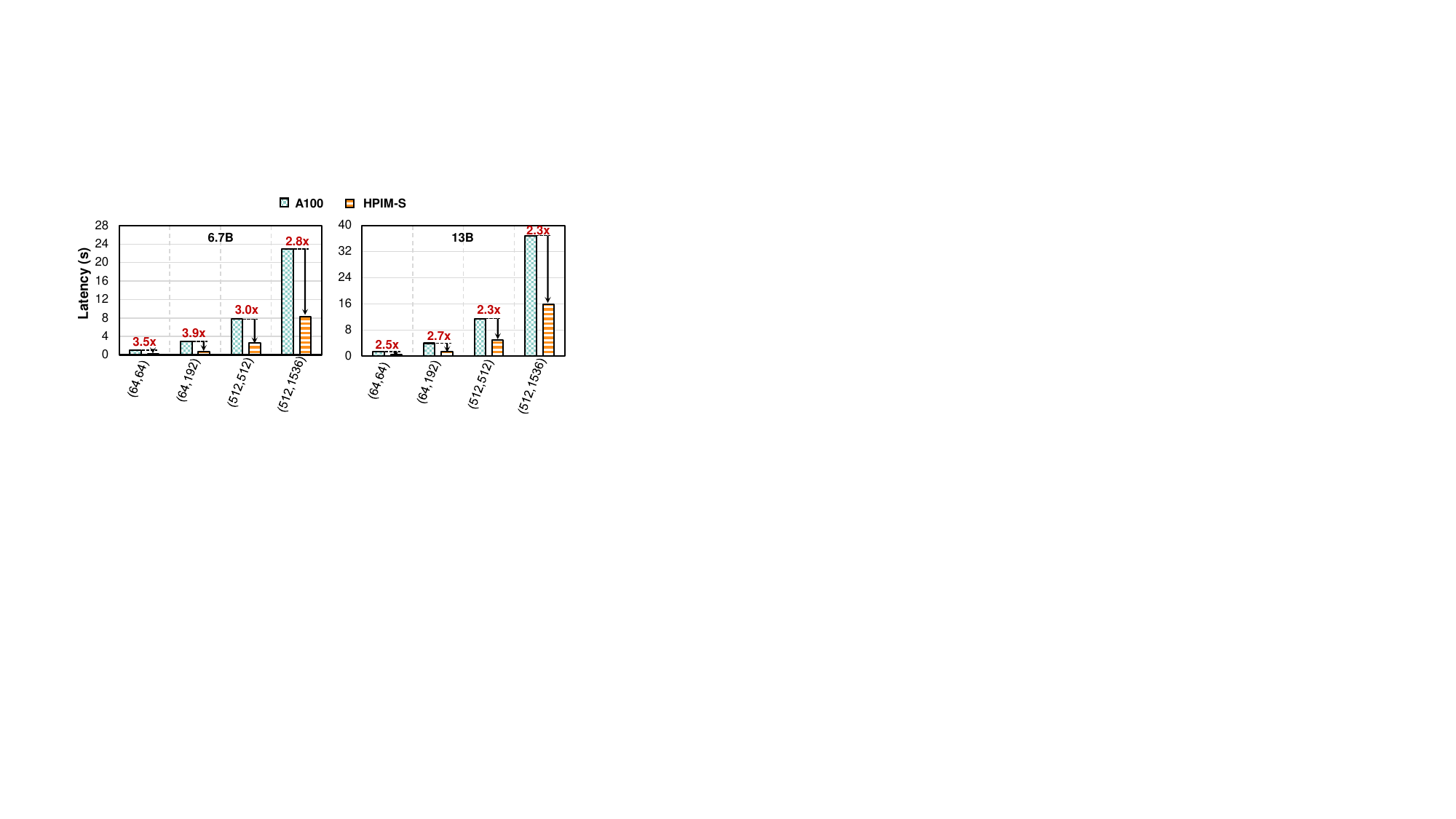}
\vspace{-8pt}
\caption{End-to-end inference latency comparison between HPIM-S and NVIDIA A100 GPU in OPT 6.7B and OPT 13B.}
\label{fig:hpim-s}
\end{figure}

\begin{figure}[t]
\centering
\includegraphics[width =0.9\linewidth]{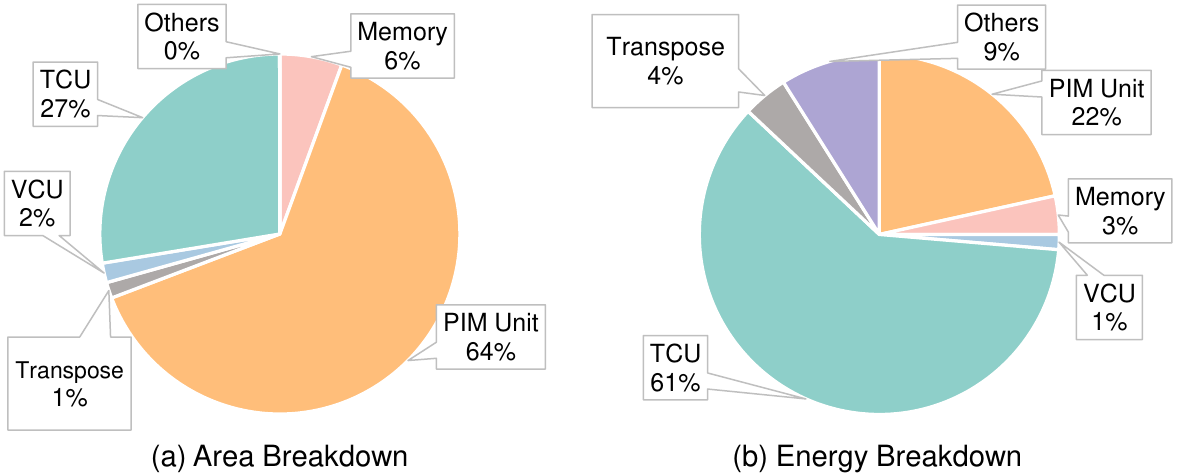}
\vspace{-4pt}
\caption{\hlblue{Area and Energy Breakdown analysis of SRAM-PIM subsystem.}}
\label{fig:area/power}
\end{figure}

\begin{figure}[t]
\centering
\includegraphics[width =0.9\linewidth]{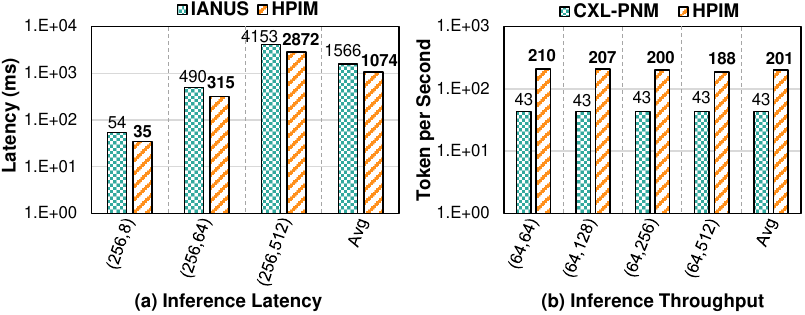}
\vspace{-4pt}
\caption{\hl{Performance comparison of HPIM against SOTA accelerators IANUS and CXL-PNM on the OPT-13B model in the decoding phase. (a) Decoding latency comparison with IANUS. (b) Decoding throughput comparison with CXL-PNM.}}
\label{fig:throughput}
\end{figure}

\hl{\textbf{Architectural Advantage Verification.} To further isolate the performance benefits of the HPIM architecture from hardware resource advantages, we introduce a scaled-down variant, HPIM-S. While the baseline HPIM offers 1.7$\times$ the DRAM bandwidth of NVIDIA A100, HPIM-S is configured with halved HBM-PIM modules.
It achieves a total DRAM bandwidth of 1638 GB/s, which is slightly below the 1935 GB/s provided by the NVIDIA A100.
Meanwhile, the peak compute throughput of HPIM-S is still substantially lower than that of the A100 GPU (312 TFLOPS).
As illustrated in Fig.~\ref{fig:hpim-s}, HPIM-S consistently achieves a significant speedup over the A100, reaching a performance improvement of up to 3.5$\times$ despite the bandwidth and throughput deficit. Notably, the performance degradation from HPIM to HPIM-S is disproportionately small compared to the 50\% bandwidth reduction. For instance, in the compute-intensive OPT-13B (512, 1536) scenario, the speedup only marginally decreases from 2.7$\times$ to 2.3$\times$. These results demonstrate that HPIM’s superior performance is fundamentally derived from its heterogeneous PIM coordination and intra-token pipelining strategies, rather than a mere escalation of hardware resources. This underscores the high architectural efficiency and scalability of the proposed memory-centric heterogeneous PIM design.}

\textbf{\hl{Evaluation of Power and Area.}}
\hblue{
Fig.~\ref{fig:area/power} presents the area breakdown of the proposed SRAM-PIM subsystem and its energy characteristics for OPT-30B inference with sequence lengths of (512, 1536). \hlblue{Fig.~\ref{fig:area/power}(a) shows the area breakdown, with a total area of 252.9 mm$^2$. The PIM Unit occupies the largest share (64\%), providing the compute density required for GEMV-intensive decoding execution.
The TCU follows with 27\% of the total area. 
The control and interconnect overhead remains negligible ($<1\%$), indicating high area efficiency for active computation.}
For power evaluation, the proposed SRAM-PIM subsystem achieves 13.5~W in the prefill stage and 1.2~W in the decoding stage. Fig.~\ref{fig:area/power}(b) shows the combined energy breakdown for the entire inference process. The TCU dominates the total energy consumption with 61\%, owing to the GEMM-intensive prefill execution, while the PIM Unit contributes 22\%, mainly from attention-related computations during decoding. The remaining modules account for only minor portions of the total energy.}

\textbf{Comparison with Other SOTA Designs.}
To demonstrate the superiority of our architecture, we benchmark HPIM on the OPT-13B model against both the IANUS and the CXL-PNM platforms. 
For a fair comparison, we maintain identical input/output token configurations as reported in prior work \cite{seo2024ianus, park2024lpddr}.
To accommodate the memory requirements of the 13B model, the IANUS system utilizes four devices interconnected via a PCIe $5.0$ $\times16$ host interface.

\hblue{As depicted in Fig.~\ref{fig:throughput}(a), HPIM consistently outperforms the IANUS system in decoding latency across all tested sequence lengths.
Since the HPIM architecture is designed to overcome the memory bottlenecks in autoregressive generation, this targeted evaluation focuses exclusively on the decoding phase to validate our core architectural innovations.
HPIM achieves up to 1.56$\times$ speedup compared to IANUS.
This performance improvement is attributed to three core architectural innovations.
First, HPIM offloads attention GEMVs to the SRAM-PIM core, effectively circumventing the resource underutilization inherent in IANUS’s NPU approach. 
Second, for weight-intensive FC layers, the HBM3-PIM subsystem leverages its massive internal bandwidth to compensate for its lower operating frequency relative to IANUS.
Third, HPIM's monolithic scale-up design entirely avoids the significant communication overhead inherent in a multi-node scale-out system like IANUS, where inter-device data synchronization over PCIe becomes a critical bottleneck.}

\hl{This trend of architectural superiority is further corroborated when comparing decoding throughput against the CXL-PNM.
Fig.~\ref{fig:throughput}(b) illustrates that HPIM delivers substantially higher tokens per second (TPS), achieving a peak of up to $4.88\times$ over CXL-PNM. 
This advantage is rooted in HPIM's true in-memory computing paradigm, which harnesses the massive internal bandwidth of HBM3 and SRAM.} 
In contrast, CXL-PNM, a near-memory solution, remains constrained by the limited bandwidth of its external CXL interconnect.
Overall, these results demonstrate the effectiveness of HPIM’s heterogeneous, fully PIM-centric architecture. 
By mitigating the memory bottleneck directly at its source, HPIM outperforms both hybrid NPU-PIM and near-memory accelerators, highlighting the significant benefits of a dedicated, memory-centric design for large language model inference.
\begin{figure}[t]
\centering
\includegraphics[width =0.9\linewidth]{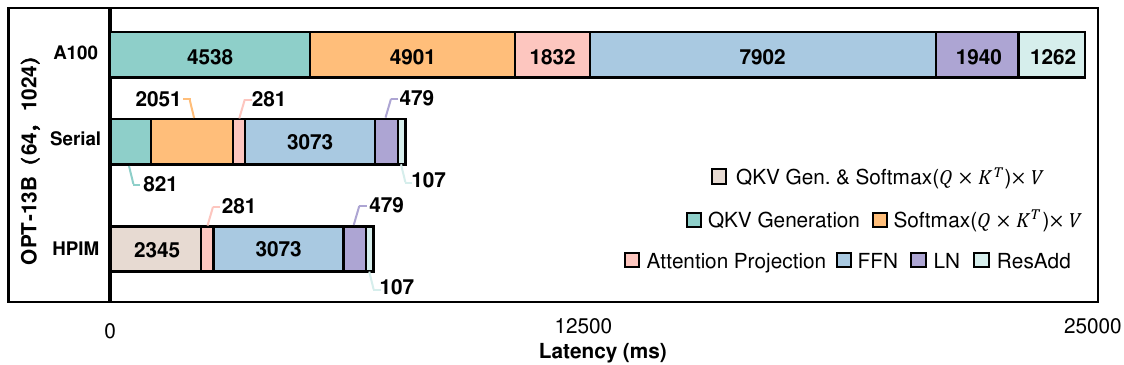}
\vspace{-8pt}
\caption{Layer-wise latency breakdown during the decoding stage for OPT-13B, evaluated on HPIM and A100 platforms.}
\label{Fig13-latency}
\end{figure}

\textbf{Layer-Wise Breakdown Analysis.}
\hl{
Fig.~\ref{Fig13-latency} illustrates the operator-level latency breakdown of OPT-13B decoding on the A100 and HPIM under serial and pipelined execution.
First, HPIM exploits heterogeneous memory characteristics for targeted acceleration. Weight-intensive GEMV operators are mapped to the HBM-PIM subsystem to leverage its massive internal bandwidth, while latency-critical attention operations are assigned to the SRAM-PIM subsystem for high-speed in-situ computation.
\hblue{Second, the comparison between the serial baseline and HPIM reveals that fine-grained cross-subsystem pipelining effectively overlaps QKV generation with attention execution, consequently shortening the decoding critical path.
While the strictly sequential dependencies of single-batch inference introduce pipeline bubbles and an effective idle rate of 55.3\% in SRAM-PIM, the HBM-PIM subsystem remains active throughout the decoding process and reaches a compute utilization of 42\%. 
This indicates that the weight-intensive GEMV path is sustained efficiently in situ on HBM-PIM rather than suffering from severe underutilization.}
Crucially, we prioritize minimizing the critical-path latency for real-time interaction over maximizing the peak utilization of every heterogeneous unit.
In summary, this combination of targeted hardware acceleration and system-level pipeline parallelism results in a comprehensive 3.56$\times$ speedup over the NVIDIA A100.
}

\begin{figure}[t]
\centering
\includegraphics[width =0.8\linewidth]{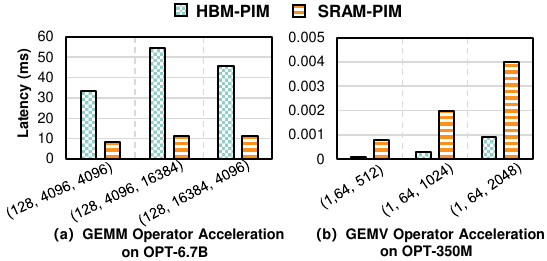}
\vspace{-6pt}
\caption{Operator-level analysis evaluated on HBM-PIM and SRAM-PIM.}
\vspace{-7pt}
\label{Fig11-operators}
\end{figure}

\begin{figure}[t]
\centering
\includegraphics[width =0.8\linewidth]{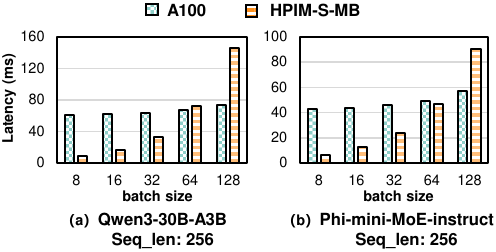}
\vspace{-6pt}
\caption{Comparison of multi-batch inference latency on A100 and HPIM-S-MB.}
\vspace{-8pt}
\label{Fig12-multi-batch}
\end{figure}
\hl{\textbf{Operator-Level Analysis.} To validate the effectiveness of our heterogeneous design, we compared the execution times for accelerating GEMM and GEMV operators on the two subsystems, respectively. First, to justify the integration of the dedicated TCU, we evaluated the performance of GEMM kernels. As illustrated in Fig.~\ref{Fig11-operators}, the SRAM-PIM TCU significantly outperforms the HBM-PIM subsystem, confirming that the high-throughput TCU is indispensable for minimizing Time-to-First-Token (TTFT). Second, regarding latency-critical Attention kernels ($d=64$), we observed a trade-off regarding data locality. For a standalone GEMV, HBM-PIM benefits from zero data transfer overhead since the data naturally resides in HBM. Although SRAM-PIM achieves faster computation latency, it requires an initial data transfer from HBM to SRAM. Consequently, due to this transfer overhead, the total latency of SRAM-PIM is higher than that of HBM-PIM for a single GEMV kernel. However, for the complete Attention operator (involving Transpose and Softmax), SRAM-PIM gains a decisive advantage. It fetches the data once and executes the complex dependency chain strictly in situ without further external access, whereas HBM-PIM would necessitate expensive round-trip data movement for these operations. This results in significantly lower end-to-end latency on SRAM-PIM, validating our mapping strategy.}

\subsection{\hblue{Multi-Batch Inference Evaluation}}
\hblue{Although HPIM is primarily optimized for single-batch LLM inference, we further evaluate it under multi-batch inference using two representative Mixture-of-Experts (MoE) models.
Thus, we introduce HPIM-S-MB, an extended configuration of HPIM-S specifically designed to support Multi-Batch (MB) inference.
As shown in Fig.~\ref{Fig12-multi-batch}, HPIM-S-MB provides clear performance gains over the A100 baseline at small batch sizes, demonstrating that its heterogeneous PIM design remains effective beyond the target single-batch scenario. 
However, this advantage diminishes as batch size increases, and HPIM-S-MB is eventually outperformed by the A100 baseline at large batch sizes.
Notably, these results are obtained with only the basic mechanisms required to support multi-batch execution, without specialized optimization for batch-level parallelism. 
A fully optimized multi-batch co-design is beyond the scope of this paper and is left for future work.}

\begin{figure}[t]
\centering
\includegraphics[width =\linewidth]{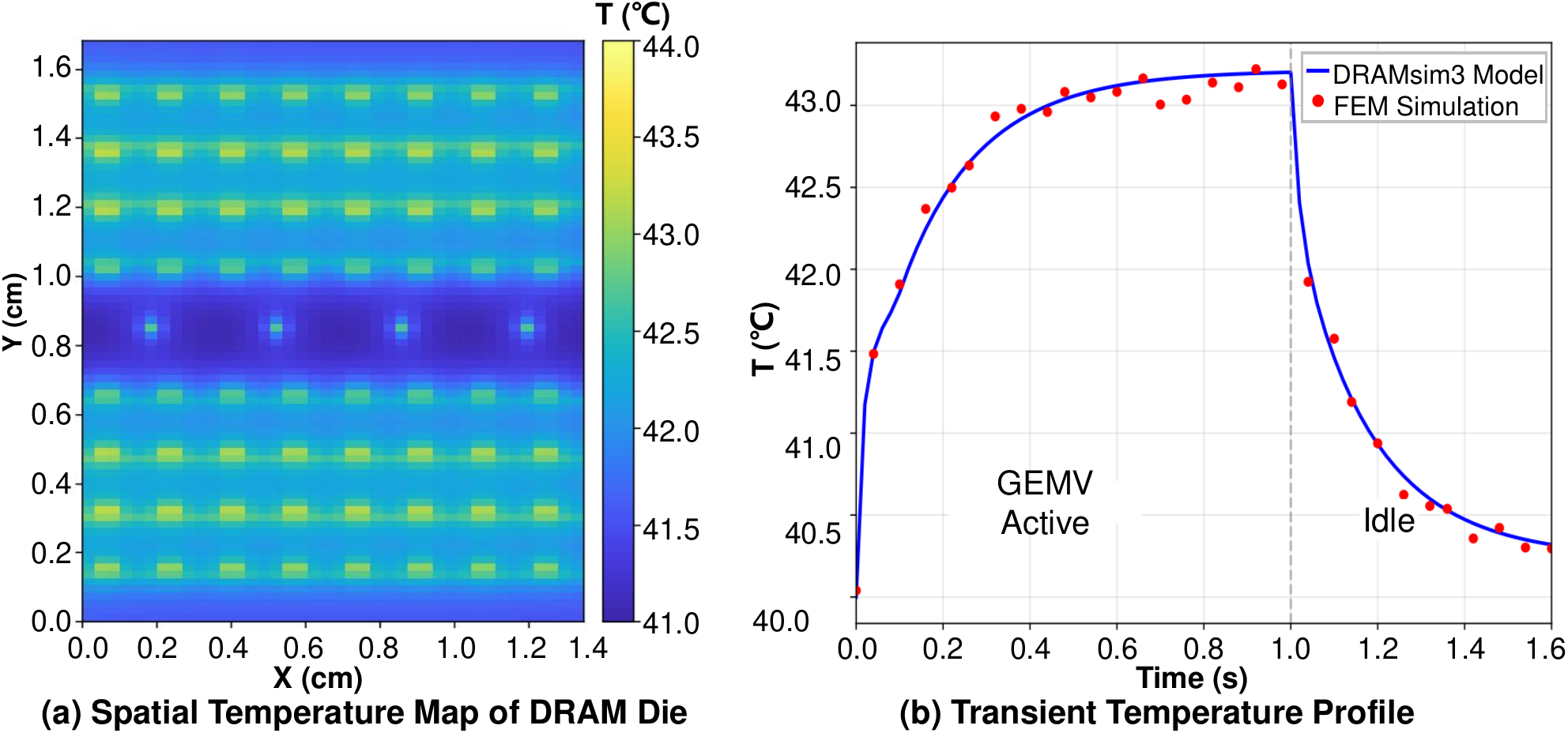}
\vspace{-6pt}
\caption{Coarse-grained thermal analysis of HPIM. (a) Spatial temperature map of DRAM die, and (b) Transient temperature profile across GEMV active and idle phases, comparing the DRAMsim3 model with FEM simulation.}
\label{Fig13-thermal}
\end{figure}
\subsection{\hlblue{Thermal Sensitivity Analysis}}
\hlblue{Thermal behavior is an important practical consideration for 3D-integrated HPIM architectures.
To estimate the thermal impact in HPIM, we conduct a coarse-grained thermal sensitivity analysis using an extended version of DRAMSim3~\cite{li2020dramsim3}. 
This analysis is intended to estimate the temperature impact range under power and placement assumptions, rather than to provide a detailed and accurate floorplan thermal analysis.
Fig.~\ref{Fig13-thermal}(a) shows the 2D temperature map of the DRAM die extracted from the full 3D HBM-stack simulation. 
The peak localized temperature within the HBM3 stack remains below 44$^\circ$C under the evaluated assumptions. 
Furthermore, Fig.~\ref{Fig13-thermal}(b) tracks the transient peak temperature profile during the GEMV Active-to-Idle phase transition.
The DRAMsim3-based results follow the same rise and decay trend as the FEM numerical results under the same power and thermal-boundary assumptions. These results suggest that the HPIM does not cause a severe temperature increase under the evaluated assumptions.
}
\section{\hl{Related Work}}\label{sec:7}

PIM accelerators mitigate the memory wall in LLM inference, yet different deployment scenarios dictate distinct architectural optimizations. 
As summarized in Tab.~\ref{Comparative}, these approaches can be broadly categorized into two main classes: 
\textbf{High-Throughput, Multi-Batch Inference Architectures.} Targeting data center scenarios, these designs prioritize system throughput to maximize the volume of processed requests. 
For example, NeuPIMs~\cite{heo2024neupims} and AttAcc~\cite{park2024attacc} both employ a heterogeneous "NPU+HBM PIM" system, leveraging pipelining and spatial parallelism across multiple requests to maximize hardware utilization and throughput.
\textbf{Low-Latency, Single-Batch Architectures.} Real-time interactive applications require minimizing the end-to-end latency of a single request. 
Heterogeneous designs, such as Flash-based accelerators~\cite{yu2024cambricon} and GDDR6-PIM systems with NPUs~\cite{seo2024ianus}, reduce data movement by combining high-capacity memory with dedicated compute engines.
Monolithic architectures, such as FIMDRAM~\cite{9365862} and TransPIM~\cite{zhou2022transpim}, focus on unlocking the potential of HBM internal bandwidth.
FIMDRAM~\cite{9365862} represents a significant milestone as a silicon-proven HBM2-PIM, demonstrating the efficacy of Bank-Level Parallelism for accelerating memory-bound GEMV kernels.
TransPIM~\cite{zhou2022transpim} proposes a software-hardware co-design for the whole LLM inference.
However, purely monolithic HBM approaches often face practical challenges regarding logic density and thermal dissipation compared to heterogeneous solutions.


\begin{table}[]
\centering
\caption{An analysis of SOTA accelerators for LLM inference.}
\label{Comparative}
\resizebox{\linewidth}{!}{
\begin{tabular}{|c|c|c|c|c|}
\hline
\textbf{Arch.} & \textbf{\begin{tabular}[c]{@{}c@{}}Subsystem \\ for GEMM\end{tabular}}            & \textbf{\begin{tabular}[c]{@{}c@{}}Subsystem \\ for GEMV\end{tabular}}    & \textbf{\begin{tabular}[c]{@{}c@{}}Inference\\ Focus\end{tabular}} & \textbf{Advantages}                                                                                                           \\ \hline
NeuPIMs~\cite{heo2024neupims}        & NPU                                                                               & HBM-PIM                                                                  & Multi                                                              & High Throughput                                                                                                               \\ \hline
AttAcc~\cite{park2024attacc}         & NPU                                                                               & HBM-PIM                                                                   & Multi                                                              & High Throughput                                                                                                               \\ \hline
IANUS~\cite{seo2024ianus}          & NPU                                                                               & GDDR6                                                                     & Single                                                             & Unified PIM-centric Design                                                                                                    \\ \hline
Cambricon-LLM~\cite{yu2024cambricon}  & NPU                                                                               & Flash                                                                     & Single                                                             & Extreme Capacity, Low Cost                                                                                                              \\ \hline
CXL-PNM~\cite{park2024lpddr}         & PE                                                                                & LPDDR5X                                                                   & Single                                                             & Scalable \& TCO-Efficient                                                                                                     \\ \hline
TransPIM~\cite{zhou2022transpim}       & N/A                                                                               & HBM-PIM                                                                   & Single                                                             & \begin{tabular}[c]{@{}c@{}}Monolithic PIM with \\ High Bandwidth\end{tabular}                                                                                             \\ \hline
FIMDRAM ~\cite{9365862}       & N/A                                                                               & HBM-PIM                                                                   & GEMV                                                            & \begin{tabular}[c]{@{}c@{}}Monolithic PIM with \\ High Parallelism \& Silicon Proven \end{tabular}                                                                                             \\ \hline
\end{tabular}
}
\end{table}

\section{Conclusion}\label{sec: Conclusion}
This paper introduces HPIM, a novel memory-centric heterogeneous Processing-In-Memory accelerator designed to address the critical performance bottlenecks in single-batch LLM inference. 
HPIM integrates an HBM-PIM subsystem for weight-intensive GEMV operations and an SRAM-PIM subsystem for latency-critical attention computations. 
This hardware-aware partitioning, combined with a deep pipeline strategy, unlocks intra-token parallelism to directly mitigate the serial dependency inherent in autoregressive decoding.
Compared to a state-of-the-art NVIDIA A100 GPU, HPIM achieves a peak speedup of up to $23.1\times$. 
\hblue{Furthermore, HPIM outperforms other state-of-the-art PIM-based accelerators, showing up to $1.56\times$ speedup over IANUS and up to $4.88\times$ higher throughput than CXL-PNM. 
These results validate that by embracing a heterogeneous, memory-centric design, HPIM provides a practical and effective solution for accelerating LLM inference.}
\section{Acknowledgement}
\hblue{The authors acknowledge the use of Gemini to enhance the readability and clarity of the text. The AI tool was used strictly for language polishing and grammatical corrections. No original ideas, concepts, or core research data were generated by the AI tool. The authors take full responsibility for the final content and integrity of this manuscript.}

{
\small
\bibliographystyle{IEEEtran}
\bibliography{ref.bib}
}


\vspace{-8mm}

\begin{IEEEbiography}[{\includegraphics[width=1in,height=1.25in,clip,keepaspectratio]{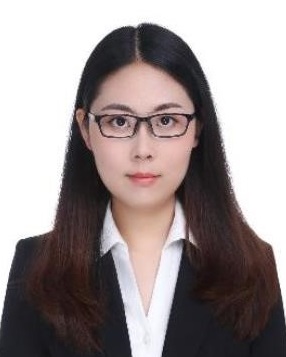}}]{Cenlin Duan}

received the B.S. degree in electronic science and technology from the University of Electronic Science and Technology of China, Chengdu, China, in 2015, the M.S. degree in software engineering from Xidian University, Xi’an, China, in 2018, and the Ph.D. degree in integrated circuit science and engineering from Beihang University, Beijing, China, in 2026. She is currently a Postdoctoral Researcher with the School of Integrated Circuit Science and Engineering, Beihang University. Her current research interests include processing-in-memory architectures and deep learning accelerators.

\end{IEEEbiography}

\vspace{-8mm}

\begin{IEEEbiography}[{\includegraphics[width=1in,height=1.25in,clip,keepaspectratio]{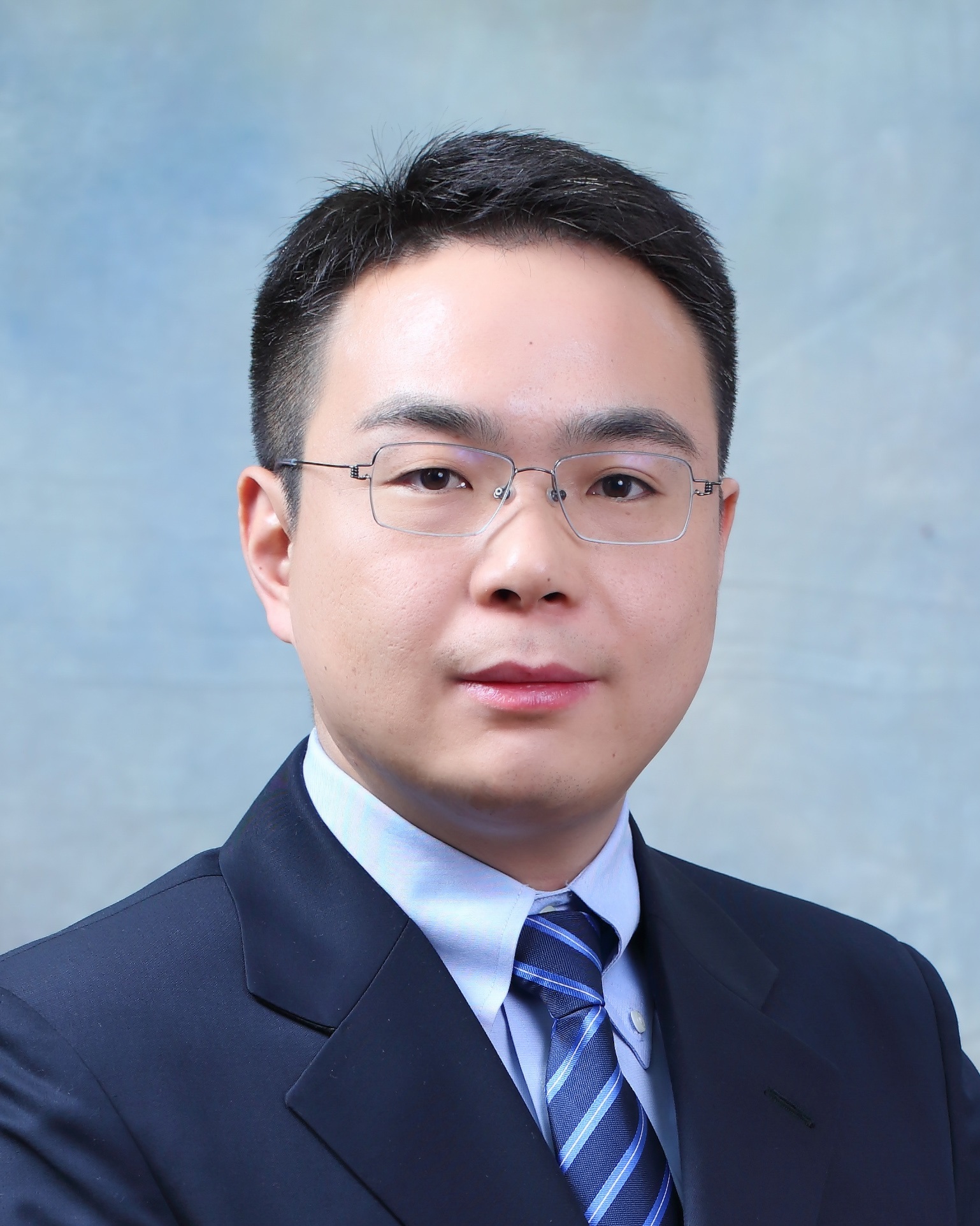}}]{Jianlei Yang}

(Senior Member, IEEE) received the B.S. degree in microelectronics from Xidian University, Xi'an, China, in 2009, and the Ph.D. degree in computer science and technology from Tsinghua University, Beijing, China, in 2014.

He is currently a Professor in Beihang University, Beijing, China, with the School of Computer Science and Engineering. From 2014 to 2016, he was a post-doctoral researcher with the Department of ECE, University of Pittsburgh, Pennsylvania, USA.
His current research interests include emerging computer architectures, hardware-software co-design, and machine learning systems. 

Dr. Yang was the recipient of the First/Second place on ACM TAU Power Grid Simulation Contest in 2011 and 2012. He was a recipient of IEEE ICCD Best Paper Award in 2013, ACM GLSVLSI Best Paper Nomination in 2015, IEEE ICESS Best Paper Award in 2017, ACM SIGKDD Best Student Paper Award in 2020.

\end{IEEEbiography}

\vspace{-8mm}

\begin{IEEEbiography}[{\includegraphics[width=1in,height=1.25in,clip,keepaspectratio]{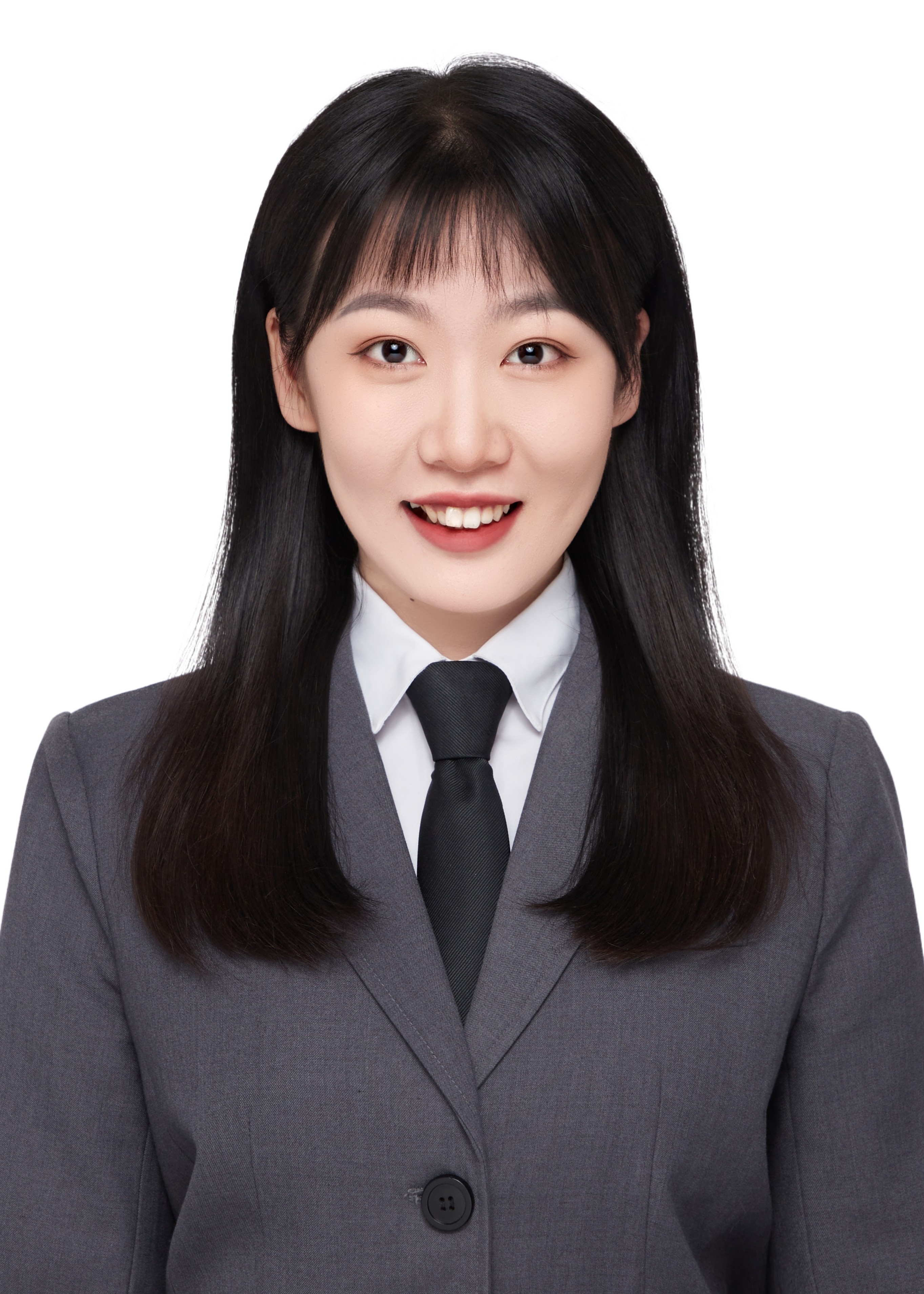}}]{Rubing Yang}

received the B.S. and M.S. degrees in computer science and technology from Beihang University, Beijing, China, in 2023 and 2026. Her research interests include computing-in-memory architectures and large language model acceleration.

\end{IEEEbiography}

\vspace{-8mm}

\begin{IEEEbiography}[{\includegraphics[width=1in,height=1.25in,clip,keepaspectratio]{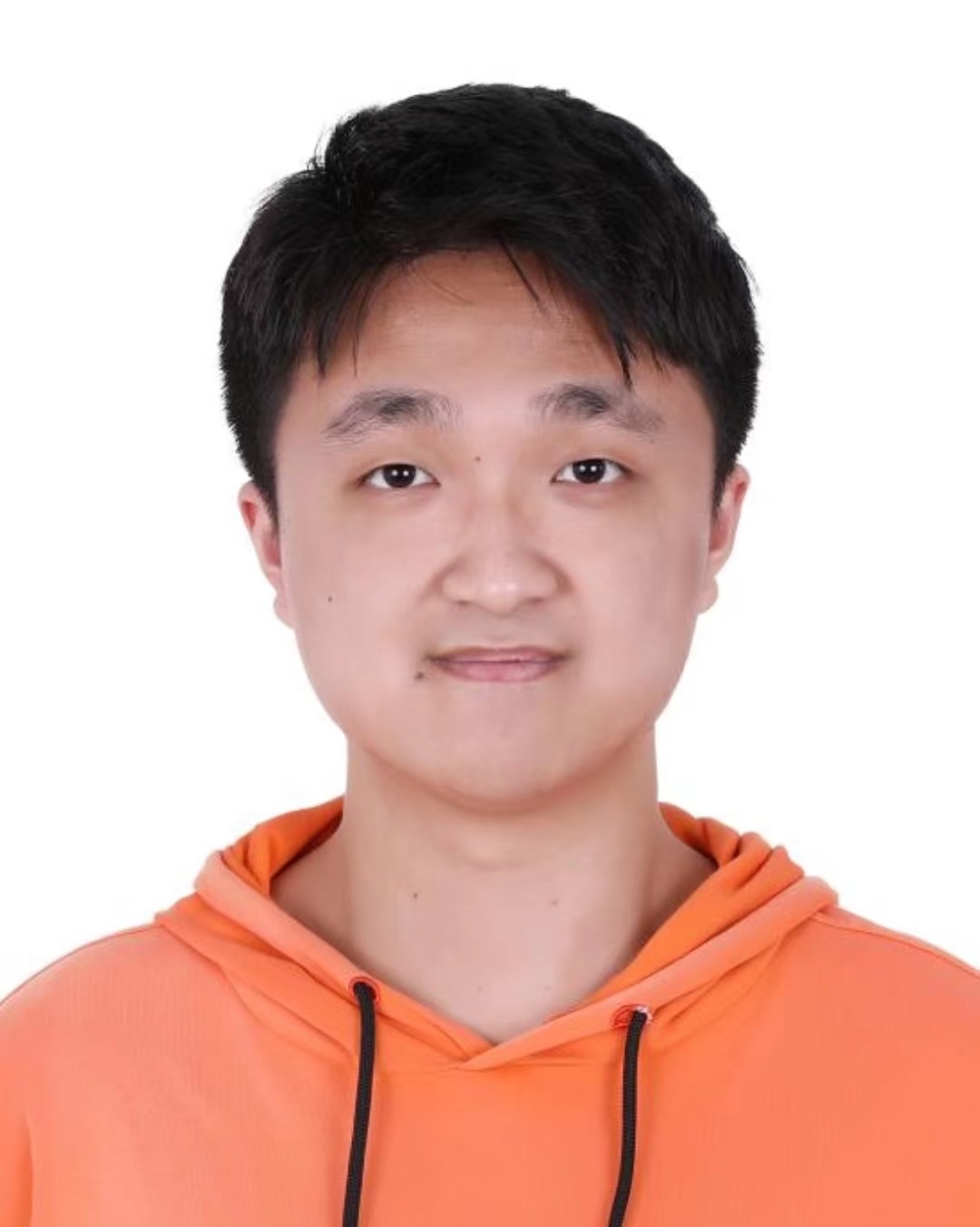}}]{Yikun Wang}

received the B.S. degree in computer science and technology from Beihang University, Beijing, China, in 2022, and the M.S. degree at the School of Computer Science and Engineering, Beihang University, China, in 2025. His current research interests include computing-in-memory architectures and deep learning accelerators.

\end{IEEEbiography}

\vspace{-8mm}

\begin{IEEEbiography}[{\includegraphics[width=1in,height=1.25in,clip,keepaspectratio]{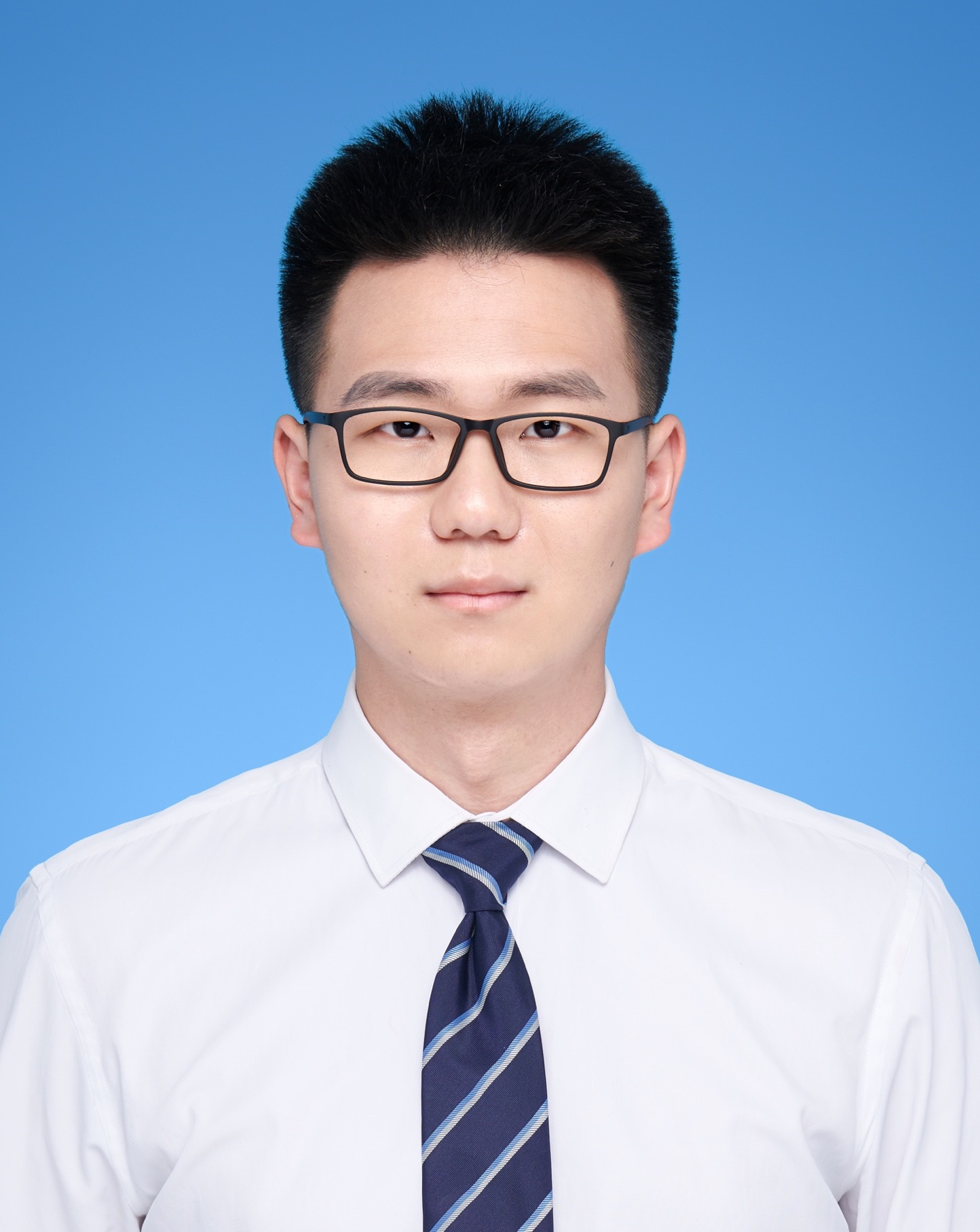}}]{Yiou Wang}

received the B.S. degree in computer science and technology from Beijing University of Technology, Beijing, China, in 2022, and the M.S. degree at the School of Computer Science and Engineering, Beihang University, China, in 2025. His current research interests include deep learning compilers.

\end{IEEEbiography}

\vspace{-8mm}

\begin{IEEEbiography}[{\includegraphics[width=1in,height=1.25in,clip,keepaspectratio]{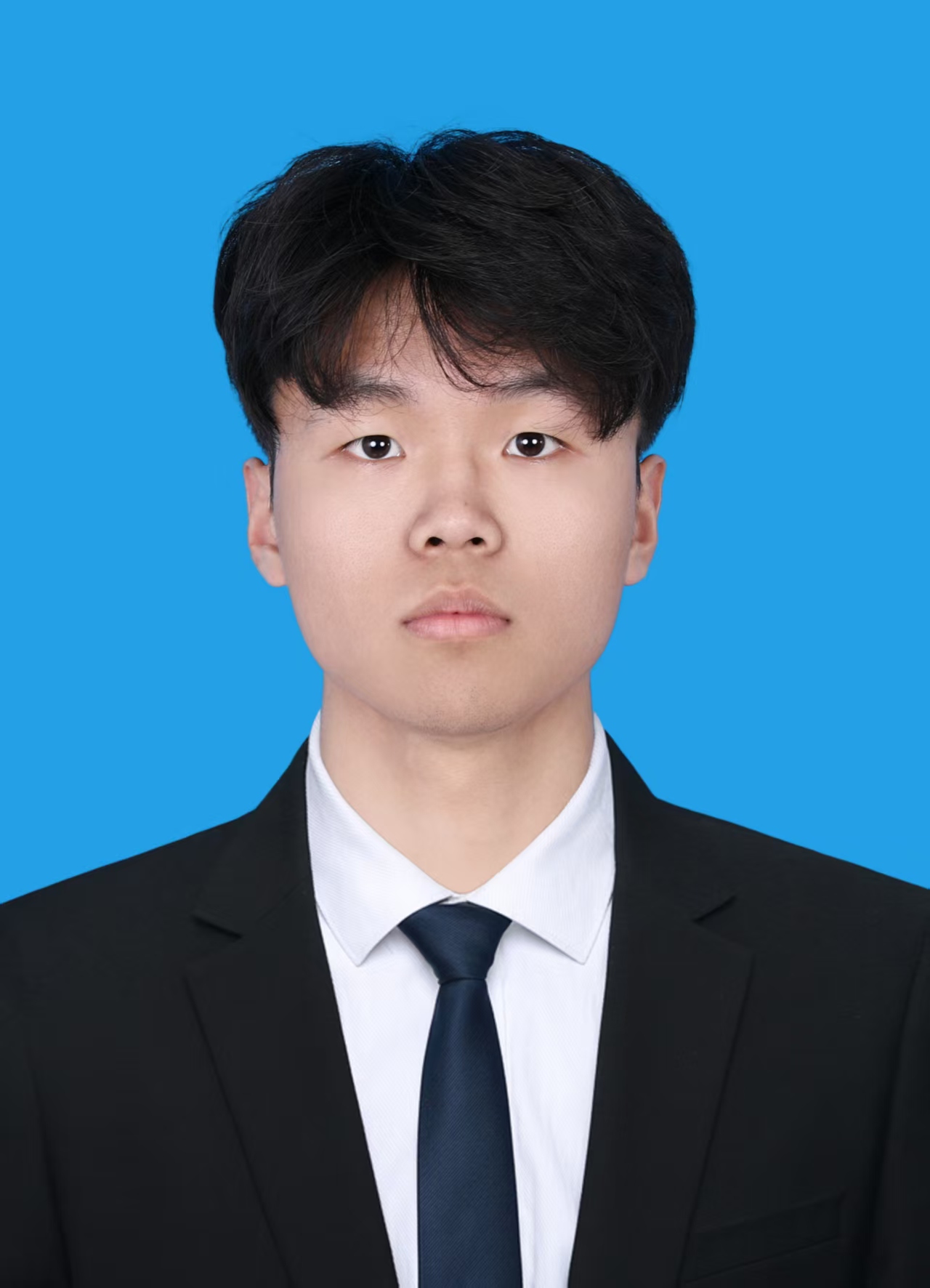}}]{Lingkun Long}

received the B.S. degree in Computer Science and Technology from Beihang University, Beijing, China, in 2024. He is currently working toward the M.S. degree in Computer Science and Technology in Beihang University, Beijing, China. His current research interests include Efficient AI and ML systems.

\end{IEEEbiography}

\vspace{-8mm}

\begin{IEEEbiography}[{\includegraphics[width=1in,height=1.25in,clip,keepaspectratio]{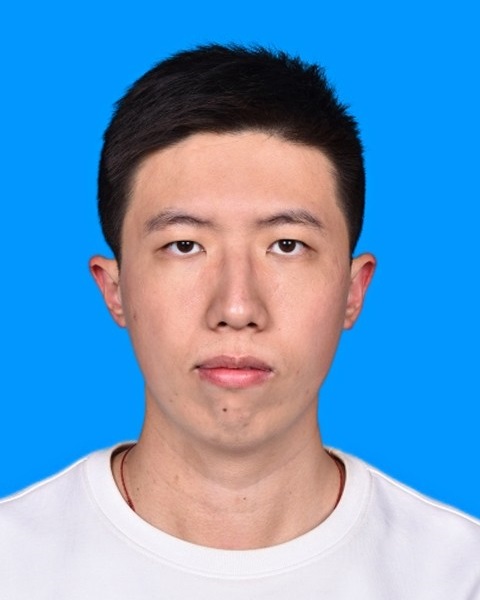}}]{Yingjie Qi}

received the B.S. degree in computer science and technology from Beihang University, Beijing, China, in 2020. He is currently pursuing the Ph.D. degree at the School of Computer Science and Engineering, Beihang University, China. His research interests include graph neural networks acceleration, compute-in-memory architectures and deep learning compilers.

\end{IEEEbiography}

\vspace{-8mm}

\begin{IEEEbiography}[{\includegraphics[width=1in,height=1.25in,clip,keepaspectratio]{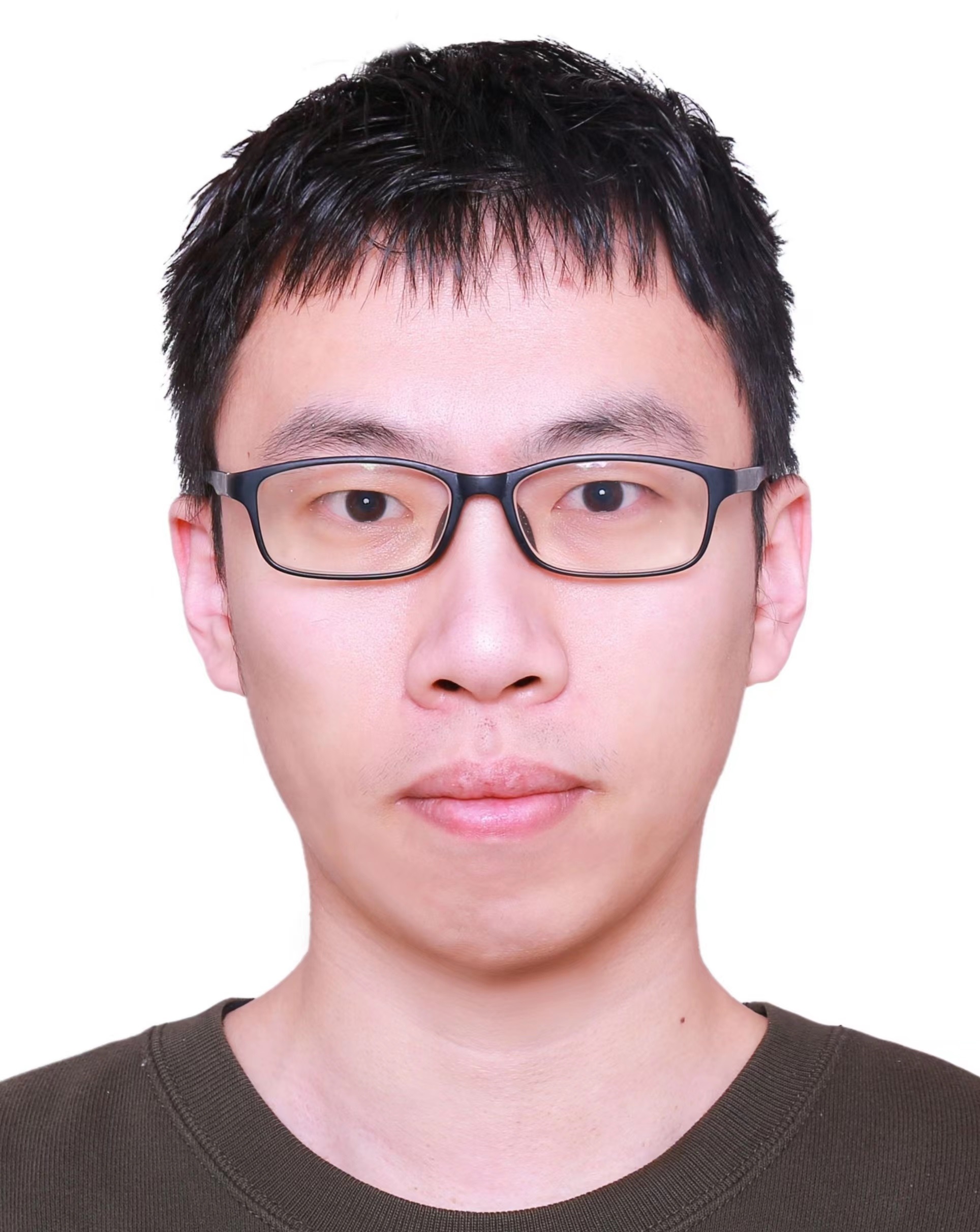}}]{Xiaolin He}

received the B.S. degree in software engineering from Beihang University, Beijing, China, in 2020. He is currently pursuing the Ph.D. degree at the School of Computer Science and Engineering, Beihang University, China. His research interests include in-memory computing architectures and compiler optimization techniques.

\end{IEEEbiography}

\vspace{-8mm}

\begin{IEEEbiography}[{\includegraphics[width=1in,height=1.25in,clip,keepaspectratio]{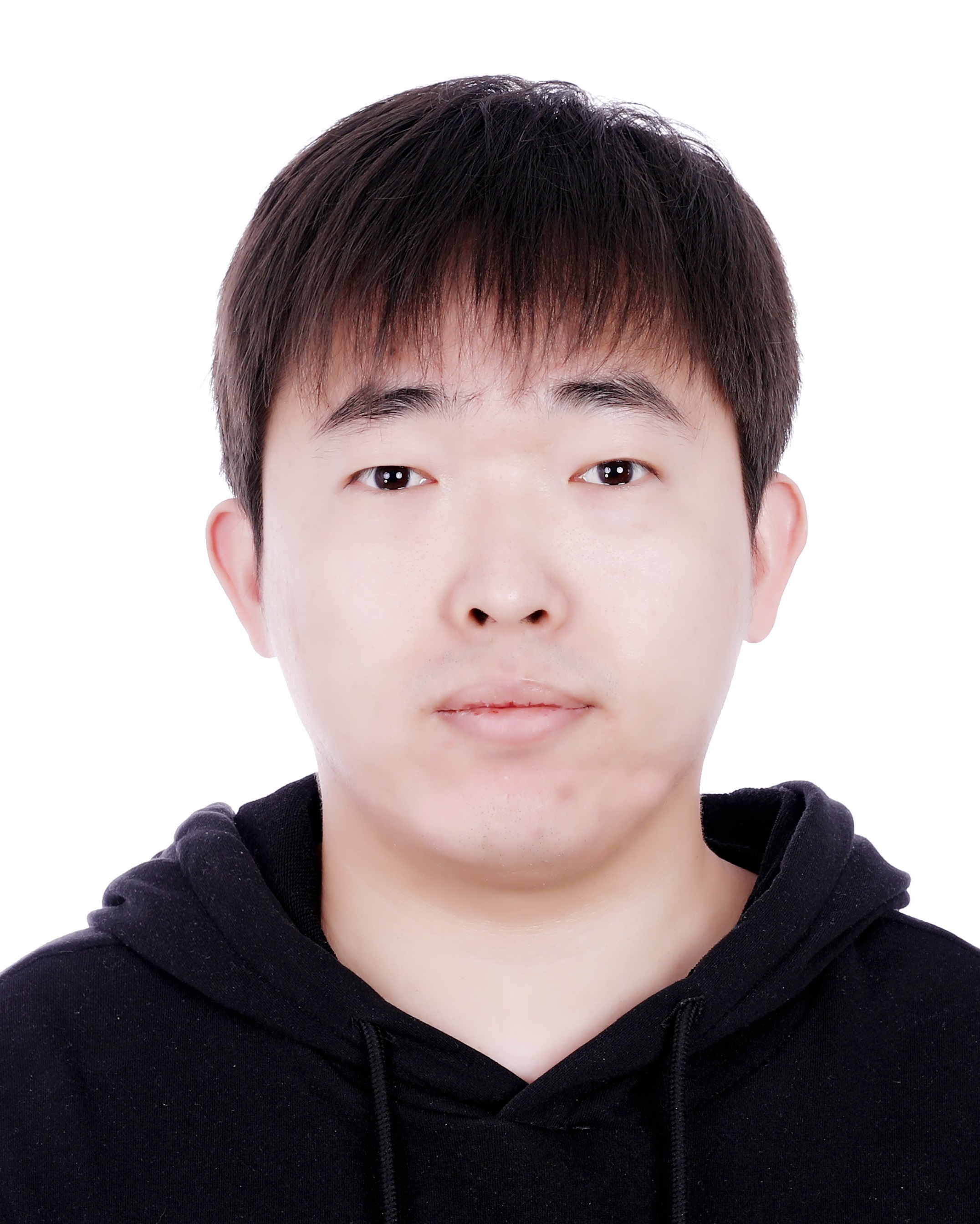}}]{Ao Zhou}

received the B.S. and M.S. degrees in software engineering from Beijing University of Technology, Beijing, China, in 2018 and 2021, respectively, and the Ph.D. degree in software engineering from Beihang University, Beijing, China, in 2025. He is currently a Postdoctoral Researcher at the School of Software, Beihang University. His research interests include GNN acceleration, computer architecture, FPGA accelerator, and heterogeneous computing. He is one of the contributors to the popular GNN computation framework PyG.

\end{IEEEbiography}

\vspace{-8mm}

\begin{IEEEbiography}[{\includegraphics[width=1in,height=1.25in,clip,keepaspectratio]{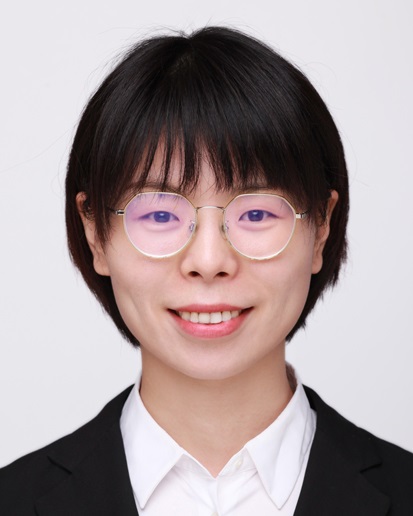}}]{Xueyan Wang}

(Member, IEEE) received the B.S. degree in computer science and technology from Shandong University, Jinan, China, in 2013, and the Ph.D. degree in computer science and technology from Tsinghua University, Beijing, China, in 2018. From 2015 to 2016, she worked as a Research Scholar at the University of Maryland, College Park, MD, USA. She is currently an Associate Professor with the School of Integrated Circuit Science and Engineering, Beihang University. She has authored more than 40 technical papers in leading journals and conferences. Her primary research interests are in the area of emerging energy-efficient computing architectures, artificial intelligence chips, and intelligent systems. 
Dr. Wang is a recipient of Young Elite Scientists Sponsorship Program by China Association for Science and Technology.

\end{IEEEbiography}

\vspace{-8mm}

\begin{IEEEbiography}[{\includegraphics[width=1in,height=1.25in,clip,keepaspectratio]{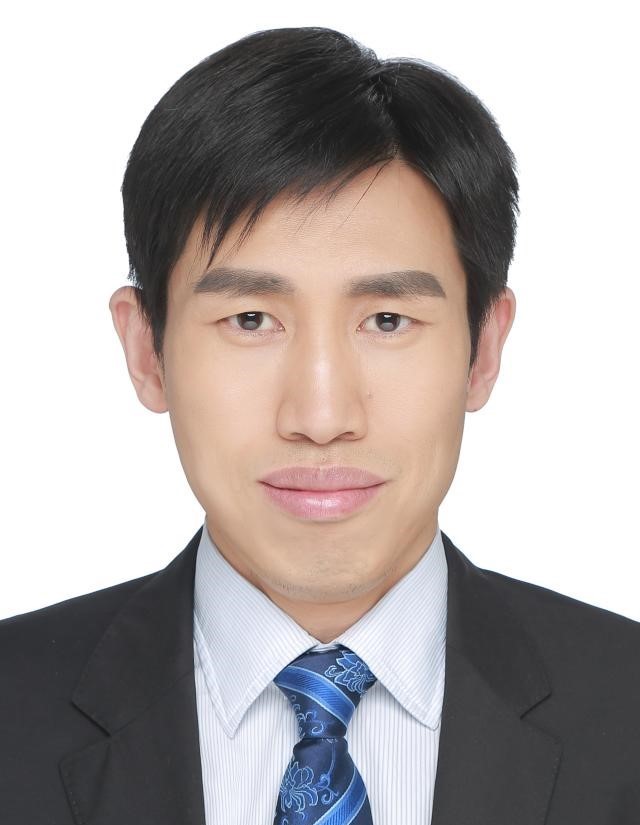}}]{Weisheng Zhao}

(Fellow, IEEE) received the Ph.D. degree in physics from the University of Paris Sud, Paris, France, in 2007.

He is currently a Professor with the School of Integrated Circuit Science and Engineering, Beihang University, Beijing, China. In 2009, he joined the French National Research Center, Paris, as a Tenured Research Scientist. Since 2014, he has been a Distinguished Professor with Beihang University. He has published more than 300 scientific articles in leading journals and conferences, such as \textit{Nature
Electronics}, \textit{Nature Communications}, \textit{Advanced Materials}, IEEE Transactions, ISCA, and DAC. His current research interests include the hybrid integration of nanodevices with CMOS circuit and new nonvolatile memory (40-nm technology node and below) like MRAM circuit and architecture design.

Prof. Zhao was the Editor-in-Chief for the {\sc{IEEE Transactions on Circuits and System I: Regular Paper}} from 2020 to 2023.

\end{IEEEbiography}


\end{document}